\documentclass[10pt,twocolumn,preprintnumbers,amsmath,amssymb,nofootinbib
,superscriptaddress]{revtex4-1}
\usepackage{graphicx,longtable,mathrsfs,color,array}
\usepackage[hidelinks]{hyperref}
\usepackage[usenames,dvipsnames]{xcolor} 
\usepackage{amssymb,amsmath,mathtools,mathrsfs,slashed} 
\usepackage{epsfig,subfigure,placeins,float} 
\usepackage{booktabs,longtable,ctable,multirow} 
\usepackage{exscale,relsize} 
\usepackage[normalem]{ulem} 
\usepackage{enumerate}
\usepackage{times, mathptmx} 
\usepackage[utf8]{inputenc}
\usepackage{color}
\usepackage{hyperref}
\usepackage{graphicx}
\usepackage{color}
\usepackage{graphicx,graphics}
\usepackage{bm}
\usepackage{tabularx}
\allowdisplaybreaks[1]

\begin{document}

\title{Kerr-(Anti-)de Sitter Black Holes: Perturbations and quasi-normal modes in the slow rotation limit}
\author{Oliver J. Tattersall}
\email{oliver.tattersall@physics.ox.ac.uk}
\affiliation{Astrophysics, University of Oxford, DWB, Keble Road, Oxford OX1 3RH, UK}
\date{Received \today; published -- 00, 0000}

\begin{abstract}
We study the perturbations of scalar, vector, and tensor fields in a slowly rotating Kerr-(Anti-)de Sitter black hole spacetime, presenting new and existing Schr{\"o}dinger style master equations for each type of perturbation up to linear order in black hole spin $a$. For each type of field we calculate analytical expressions for the fundamental quasi-normal mode frequencies. These frequencies are compared to existing results for Schwarzschild-de Sitter, slowly rotating Kerr, and slowly rotating Kerr-de Sitter black holes. In all cases good agreement is found between the analytic expressions and those frequencies calculated numerically. In addition, the axial and polar gravitational frequencies are shown to be isospectral to linear order in $a$ for all cases other than for \textit{both} non-zero $a$ and $\Lambda$.
\end{abstract}
\keywords{Black holes, Perturbations, Gravitational Waves}

\maketitle
\section{Introduction}\label{introduction}

Black holes are among the most captivating aspects of Einstein's Theory of General Relativity (GR) \cite{Einstein:1916vd,Schwarzschild:1916uq,Finkelstein:1958zz,Kerr:1963ud,Misner:1974qy}, and their properties have been studied extensively since the dawn of GR in the early 20th century. Of great interest to physicists and mathematicians alike is the response of black holes to perturbations. Notably, perturbed black holes `ring', emitting gravitational waves at a characteristic set of frequencies known as the quasi-normal mode (QNM) frequencies \cite{1975RSPSA.343..289C,0264-9381-16-12-201,Kokkotas:1999bd,Berti:2009kk,Konoplya:2011qq}. 

These QNM frequencies are dependent on the background properties of the black hole (e.g. mass or angular momentum), acting like a `fingerprint' for a given black hole. Furthermore, the presence of a cosmological constant, or a modification to the theory of gravity itself, can and will affect the spectrum of frequencies that a black hole will emit gravitational waves at. Thus studying the QNM frequencies of black holes (and other fields propagating on the black hole spacetime) provides a window from which to observe not only the properties of the black hole itself, but also of the wider universe and indeed of the fundamental laws governing gravity \cite{Dreyer:2003bv,Berti:2005ys,Berti:2015itd,2018PhRvD..97d4021T,Berti:2018vdi,Barack:2018yly,Brito:2018rfr}. 

From an observational point of view, given the dawn of the gravitational wave era of astronomy (with multiple direct observations of gravitational waves from mergers of highly compact objects, i.e. black holes or neutron stars, having now been made by advanced LIGO and VIRGO \cite{2016PhRvL.116f1102A,2016PhRvL.116x1103A,2017PhRvL.118v1101A,Abbott:2017oio,PhysRevLett.119.161101}), determining and detecting the QNM frequencies of the remnant black holes left perturbed after the merger of compact objects is an interesting and important area of study. 

In this paper, we will study the responses a variety of fields to linear perturbations on a black hole spacetime, and present analytical expressions for the QNM frequencies that each type of field characteristically `rings' at. The black holes we will consider will possess angular momentum and be embedded in a universe with a cosmological constant that can be positive or negative (i.e. the spacetime will be either asymptotically de Sitter or Anti-de Sitter). For the case of a positive cosmological constant, the black holes studied here will represent the kind of astrophysical black holes that we expect to observe in our universe (assuming the $\Lambda$CDM paradigm of cosmology \cite{Akrami:2018vks}). For a negative cosmological constant, on the other hand, the AdS/CFT correspondence provides an interesting motivation to study the QNMs of asymptotically Anti-de Sitter black holes as a method of gaining insight into certain conformal quantum field theories \cite{Maldacena:1997re,Nunez:2003eq,Son:2007vk,Hartnoll:2009sz,Herzog:2009xv}.

\textit{Summary}:  In section \ref{background}, we will present the background spacetime of the black holes that are to be studied in this work. In section \ref{perturbations}, we will review aspects of black hole perturbation theory before presenting second order Schr{\"o}dinger-style master equations for perturbations of massive scalar (spin $s=0$), massive vector ($s=-1$), and massless tensor ($s=-2$) fields. Some of the master equations presented are known from the literature, with others (to the author's best knowledge) being new results. In section \ref{qnmsection} we will then present analytic expressions for the QNM frequencies that satisfy each of the master equations present in section \ref{perturbations}, and compare these analytic expressions to previously obtained numerical results. Finally, in section \ref{conclusion}, we will discuss the results presented and make some concluding remarks. Throughout we will use units such that $G=c=1$.

\section{Background}\label{background}

The background spacetime that we will be concerned with in this work is that of a \textit{slowly} rotating black hole in a universe with a cosmological constant $\Lambda$. The black hole spacetime is described by the Kerr-(Anti)-de Sitter (henceforth referred to as Kerr-(A)dS) solution which, to linear order in dimensionless black hole spin $a$, is given in Boyer-Lindquist coordinates by \cite{1973blho.conf...57C}:
\begin{align}
ds^2=\bar{g}_{\mu\nu}dx^\mu dx^\nu=&-F(r)dt^2+F^{-1}(r)dr^2+r^2d\Omega^2\nonumber\\
&-2aM\left(\frac{2M}{r}+\frac{\Lambda}{3}r^2\right)\sin^2\theta dtd\phi\label{backgroundg}
\end{align}
with $d\Omega^2$ being the metric on the surface of the unit 2-sphere, $M$ the black hole mass, and we assume $|a|\ll1$.  The metric function $F(r)$ is given by:
\begin{align}
F(r)=1-\frac{2M}{r}-\frac{\Lambda}{3}r^2.
\end{align}
The spacetime is asymptotically de Sitter (thus describing a Kerr-dS black hole) for $\Lambda>0\;$; for $\Lambda<0$, we have a Kerr-AdS black hole with an asymptotically Anti-de Sitter spacetime.


\section{Perturbation Master Equations}\label{perturbations}

When considering perturbations on a spherically symmetric background spacetime, it is standard to decompose the perturbed fields into spherical harmonics to factor out the angular dependence of the perturbation \cite{1975RSPSA.343..289C,0264-9381-16-12-201,Kokkotas:1999bd,Berti:2009kk,Martel:2005ir,Ripley:2017kqg}. Using $Y^{\ell m}_{\mu_1...\mu_n}(\theta,\phi)$ to schematically represent the appropriate choice of scalar, vector, or tensor spherical harmonics depending on the perturbed field $f$ in question:
\begin{align}
f_{\mu_1...\mu_n}(t,r,\theta,\phi)=\sum_{\ell,m}f^{\ell m}(r,t)Y^{\ell m}_{\mu_1...\mu_n}(\theta,\phi).
\end{align}

About a spherically symmetric background, perturbations of different polarity (either axial/odd or polar/even) decouple from each other, greatly simplifying the analysis of the equations of motion for the perturbed fields. In addition perturbations of different $\ell$ decouple. The result is that, if we further assume a harmonic time dependence of the form $e^{-i\omega t}$, the equations of motion can often be cast into homogeneous Schr{\"o}dinger style second order differential equations for some unknown function of $r$ representing the perturbation:
\begin{align}
\left[\frac{d^2}{dr^2}+\left(\omega_{\ell m}^2-V_{\ell m}(r)\right)\right]\mathcal{F}^{\ell m}(r)=0.
\end{align}
The $\omega_{\ell m}$ are the characteristic QNM frequencies associated with the perturbation, as mentioned in section \ref{introduction}. Detailed reviews on QNMs can be found in \cite{1975RSPSA.343..289C,0264-9381-16-12-201,Kokkotas:1999bd,Berti:2009kk}. The fact that the equations of motion governing the perturbations can often be cast into a single Schr{\"o}dinger style second order differential equation is useful, with the techniques of quantum mechanics and time-independent scattering theory being available to the modern physicist to analyse such an equation (see, for example, the Appendix of \cite{Kokkotas:1999bd}). 

Unfortunately the decoupling of perturbations of different polarity and $\ell$ no longer occurs when the background is not spherically symmetric (e.g. in axisymmetric spacetimes such as the Kerr family of black holes \cite{Teukolsky:2014vca}), and thus the task of simplifying the equations of motion governing the perturbations is greatly complicated. For example, for a Kerr black hole one has to solve the more complex Teukolsky equation \cite{1973ApJ...185..635T} to find the QNM frequencies, rather than the simpler Regge-Wheeler \cite{Regge:1957td} or Zerilli \cite{Zerilli:1970se} equations that one calculates for Schwarzschild black holes. 

In \cite{2012PhRvD..86j4017P}, however, it was shown that in slowly rotating backgrounds, where the `breaking' of spherical symmetry is controlled by the dimensionless black hole spin $a$ (and terms $O(a^2)$ are neglected), it is sufficient to continue to use spherical harmonics and to treat perturbations of different polarity and $\ell$ as completely decoupled. This technique yields equations of motion that are sufficient to determine the QNM frequency spectrum of the system accurately to linear order in $a$ (and has been utilised in, for example, \cite{Pani:2013pma,Pani:2013wsa,Brito:2013wya}). The benefit of this approach is that one can continue to exploit the useful properties of spherical harmonics and often still arrive at simple second order equations of motion for the perturbed fields, whilst now including the effects of (slow) rotation. 

In the following sections we will make use of the above technique for perturbations in slowly rotating backgrounds to derive Schr{\"o}dinger style master equations for various type of perturbations on a slowly rotating Kerr-(A)dS background, keeping terms linear in black hole spin and neglecting terms $O(a^2)$ and higher. This will allow us in Section \ref{qnmsection} to determine the QNM frequency spectra for each type of perturbation to linear order in $a$. From now on we will also suppress spherical harmonic indices so as not to clutter notation, with each equation assumed to hold for a given $(\ell,m)$.


\subsection{Massive Scalar field}

First we consider a massive test scalar field $\Phi$ (such that $\Phi$ does not contribute to the background energy momentum and thus does not affect the background spacetime) propagating on the slowly rotating Kerr-(A)dS background given by eq.~(\ref{backgroundg}). Such a field obeys the massive Klein-Gordon equation:
\begin{align}
\Box\Phi=\mu^2\Phi.
\end{align}
In \cite{2012PhRvD..86j4017P,Pani:2013pma} it was shown that for scalar perturbations such that:
\begin{align}
\Phi=\sum_{\ell m}\frac{\varphi_{\ell m}(r)}{r}e^{-i\omega t}Y^{\ell m} (\theta,\phi),
\end{align}
the massive Klein-Gordon equation (linearised to first order in black hole spin $a$) takes the form:
\begin{widetext}
\begin{align}
\left[\frac{d^2}{dr_\ast^2}+\left(\omega^2-\frac{2amM\omega}{r^2}\left(\frac{2M}{r}+\frac{\Lambda}{3} r^2\right)-F(r)\left(\frac{l(l+1)}{r^2}+\frac{2M}{r^3}-\frac{2\Lambda}{3}+\mu^2\right)\right)\right]\varphi=0\label{scalareq}.
\end{align}
\end{widetext}

We see that the effective potential of eq.~(\ref{scalareq}) is modified from the usual spin zero Regge Wheeler equation \cite{1975RSPSA.343..289C,0264-9381-16-12-201,Kokkotas:1999bd,Berti:2009kk} through $\Lambda$ and $a$. Eq.~(\ref{scalareq}) is, nonetheless, still in the generic form of a Schr{\"o}dinger style wave equation. 


\subsection{Massive Vector field}

We now consider a massive vector field propagating on the black hole background. We again assume that the test field does not contribute to the background energy momentum and thus does not affect the background spacetime.

Whilst vector fields can have both axial and polar parity components of their perturbations, it was shown in \cite{2012PhRvD..86j4017P} that the polar perturbations for a \textit{massive} vector field on a slowly rotating background cannot be reconciled into a single Schr{\"o}dinger style wave equation. Thus in this paper we will only consider axial parity perturbations for simplicity. It is worth noting that, in the case of a massless vector field (i.e. electromagnetic perturbations), the equations for axial and polar perturbations coincide. 

In \cite{2012PhRvD..86j4017P} the master equation governing the axial component of a massive vector perturbation is given for a generic slowly rotating background. For the slowly rotating Kerr-AdS background that we are concerned with, the perturbation $A(r)$ satisfies:
 \begin{widetext}
\begin{align}
\left[\frac{d^2}{dr_\ast^2}+\left(\omega^2-\frac{2amM\omega}{r^2}\left(\frac{2M}{r}+\frac{\Lambda}{3} r^2\right)-F(r)\left(\frac{l(l+1)}{r^2}+\mu^2\right)\right)\right]A=0,\label{procaeq}
\end{align}
\end{widetext}
where $\mu$ is the vector field mass. In the case that $a=0$, eq.~(\ref{procaeq}) agrees with the master equations derived in \cite{Cardoso:2001bb}.


\subsection{Gravitational field}

Finally, we consider perturbations to the black hole spacetime itself, such that the metric $g_{\mu\nu}$ can be decomposed into a background part $\bar{g}$ and a perturbation $h$:
\begin{align}
g_{\mu\nu}=\bar{g}_{\mu\nu}+h_{\mu\nu},
\end{align}
where $\bar{g}_{\mu\nu}$ is given by eq.~(\ref{backgroundg}). The perturbed Einstein equations then read
\begin{align}
\delta R_{\mu\nu} + \Lambda h_{\mu\nu} = 0
\end{align}
with $\delta R_{\mu\nu}$ representing the Ricci tensor expanded to linear order in the metric perturbation $h_{\mu\nu}$.

For the gravitational perturbations we adopt the Regge-Wheeler gauge and decompose the metric perturbation into tensor spherical harmonics, with the tensor perturbation $h_{\mu\nu}$ having both axial and polar parity perturbations \cite{Regge:1957td,Rezzolla:2003ua}:
\begin{widetext}
 \begin{align}
 h_{\mu\nu,\ell m}^{ax}=&
 \begin{pmatrix}
 0&0&h_0(r)B^{\ell m}_\theta&h_0(r)B^{\ell m}_\phi\\
 0&0&h_1(r)B^{\ell m}_\theta&h_1(r)B^{\ell m}_\phi\\
 sym&sym&0&0\\
 sym&sym&0&0
 \end{pmatrix}e^{-i\omega_{\ell m} t},
 \label{hodd}\\
 h_{\mu\nu,\ell m}^{p}=&
 \begin{pmatrix}
 H_0(r)F(r)&H_1(r)&0&0\\
 sym&H_2(r)F(r)^{-1}&0&0\\
 0&0&K(r)r^2&0\\
 0&0&0&K(r)r^2\sin\theta
 \end{pmatrix}Y^{\ell m}e^{-i\omega_{\ell m} t},
 \label{heven}
\end{align}
\end{widetext}
where $sym$ indicates a symmetric entry, $B^{\ell m}_\mu$ is the axial parity vector spherical harmonic and $Y^{\ell m}$ is the standard scalar spherical harmonic, as described in \cite{Martel:2005ir,Ripley:2017kqg}.

Once again we can treat perturbations of different parity separately in order to study the QNM spectrum to linear order in $a$ \cite{2012PhRvD..86j4017P}. 

\subsubsection{Axial sector}

For the axial sector, we can define a function $Q(r)$ in terms of the perturbation fields $h_i$ which satisfies the following Schr{\"o}dinger style master equation:
\begin{widetext}
\begin{align}
\left[\frac{d^2}{dr_\ast^2}+\left(\omega^2-\frac{2amM\omega}{r^2}\left(\frac{2M}{r}+\frac{\Lambda}{3}r^2\right)-F(r)\left(\frac{l(l+1)}{r^2}-\frac{6M}{r^3}+am\frac{24M^2(3r-7M-2r^3\Lambda/3)}{l(l+1)r^6\omega}\right)\right)\right]Q=0.\label{gravaxial}
\end{align}
\end{widetext}
For $\Lambda=0$, eq.~(\ref{gravaxial}) agrees with the Schr{\"o}dinger style equation for axial gravitational perturbations of a slowly rotating Kerr black hole given in \cite{Pani:2013pma}. With $a=0$, eq.~(\ref{gravaxial}) agrees with the result of \cite{Cardoso:2001bb} for the axial perturbations of a Schwarzschild-(A)dS black hole. With $a=\Lambda=0$, we recover the familiar Regge-Wheeler equation \cite{Regge:1957td}.

\subsubsection{Polar sector}

For the polar sector, we can define a function $Z(r)$ in terms of the perturbation fields $H_i$, and $K$ which obeys the following Schr{\"o}dinger style master equation:
\begin{widetext}
\begin{align}
\left[\frac{d^2}{dr_\ast^2}+\left(\omega^2-\frac{2amM\omega}{r^2}\left(\frac{2M}{r}+\frac{\Lambda}{3}r^2\right)-F(r)\left(V_Z^{(0)}+amV^{(1)}_Z\right)\right)\right]Z=0\label{gravpolar}
\end{align}
\end{widetext}
where $V_Z^{(0)}$ is the familiar Zerilli-like potential for Schwarzschild-(A)dS perturbations given by:
\begin{align}
V_Z^{(0)}=\frac{2}{r^3}\frac{9M^3+3c^2Mr^2+c^2(1+c)r^3+3M^2(3cr-r^3\Lambda)}{(3M+cr)^2}
\end{align}
with $c=\frac{1}{2}[l(l+1)-2]$ \cite{Cardoso:2001bb}. $V^{(1)}_Z(r)$ is the $O(a)$ correction to the potential given in Appendix \ref{appendixpolar}. For $\Lambda=0$, i.e. when considering a slowly rotating Kerr black hole, $V^{(1)}_Z(r)$ agrees with the linear in spin correction to the polar potential given in \cite{Pani:2013pma}. With $a=\Lambda=0$, we recover the familiar Zerilli equation \cite{Zerilli:1970se}.

\section{Quasinormal modes}\label{qnmsection}

\subsection{Analytical Expressions}\label{analyticsec}

As a complement to other methods of calculating the QNM frequencies $\omega$ that satisfy the equations presented in section \ref{perturbations} \cite{Kokkotas:1999bd,0264-9381-16-12-201,Berti:2009kk,1985RSPSA.402..285L,Cho:2011sf}, we will present analytic expressions for the QNM frequencies calculated via the method developed in \cite{2009CQGra..26v5003D}. We direct the reader to \cite{2009CQGra..26v5003D} for the details of the method, with the important result being that the $\omega$ for each perturbation master equation can be expressed as a sum over inverse powers of $L=\ell+1/2$, with $\ell$ being the multi-polar spherical harmonic index:
\begin{align}
\omega=\sum_{k=-1}^{k=\infty}\omega_k \;L^{-k}.\label{expansion}
\end{align}

In appendix \ref{coeff} we present the $\omega_k$ that satisfy eqs.~(\ref{scalareq}), ~(\ref{procaeq}), ~(\ref{gravaxial}), and ~(\ref{gravpolar}) for the fundamental $n=0$ mode (with $n$ being the overtone index). In principle one can calculate the $\omega_k$ for arbitrary $n$, but for simplicity's sake we focus on the fundamental modes in this work. In each case we have calculated the first eight terms in the expansion, i.e. to $O(L^{-6})$. One can straightforwardly calculate terms to higher order in inverse powers of $L$ through the use of a computer algebra package. We will see in the following section that the QNM frequencies calculated via this analytic expansion method give results in very strong agreement with those calculated via, for example, 6th order WKB methods. 

In the limit that $\ell\rightarrow\infty$ the QNM frequencies, irrespective of field spin or mass, are given by:
\begin{align}
M\omega_{\ell m}=\;&\frac{\sqrt{1-9\Lambda M^2}}{6\sqrt{3}}\left(2\ell+1-i\right)+am\left(\frac{2}{27}+\frac{\Lambda M^2}{3}\right)\nonumber\\
&\;+O(\ell^{-1}).
\end{align}

A result of interest from calculating the $\omega_k$ that satisfy eq.~(\ref{gravaxial}) and (\ref{gravpolar}) analytically is that the axial and polar gravitational frequencies are isospectral to $O(L^{-3})$, with any differences between the $\omega_k$ at higher orders in $1/L$ being proportional to both $a$ and $\Lambda$ (see appendix \ref{polarcoeff}). Thus, to linear order in $a$, it is only in the case of non-zero black hole spin \textit{and} non-zero cosmological constant that the spectra of the axial and polar perturbations split. The isospectrality of the gravitational modes for a Schwarzschild black hole is well known \cite{1975RSPSA.343..289C,0264-9381-16-12-201,Kokkotas:1999bd}, with the isospectrality of Schwarzschild-dS and slowly rotating Kerr-Newman black holes having been observed numerically in \cite{Zhidenko:2003wq,Pani:2013wsa}.

It is worth noting that the expansion technique of \cite{2009CQGra..26v5003D} is designed for use in spherically symmetric background spacetimes; indeed we will see in the following section that the results for Schwarzschild-de-Sitter black holes are indeed highly accurate. In \cite{Tattersall:2018nve}, however, the expansion was nevertheless used to find the QNM frequencies of a massive scalar field on a slowly rotating Kerr background (i.e. the $\omega$ satisfying eq.~(\ref{scalareq}) with $\Lambda=0$). Good agreement was again found between those frequencies calculated numerically and those using the analytic expansion method, despite the background spacetime no longer being spherically symmetric. 

In the following section we will calculate the QNM frequencies satisfying each of eqs.~(\ref{scalareq}), ~(\ref{procaeq}), ~(\ref{gravaxial}), and ~(\ref{gravpolar}) for non-zero $a$ (i.e. on non-spherically symmetric, slowly rotating backgrounds) and investigate how well an approximation the $\omega_k$ presented in appendix \ref{coeff} provide for the QNM frequencies at linear order in black hole spin. In fact, we will see that for $a\ll 1$, the frequencies calculated in this paper provide a very good approximation to those calculated in the literature.

Furthermore, as explained in \cite{2009CQGra..26v5003D}, the analytic expansion method for calculating QNM frequencies utilised here is not applicable to asymptotically AdS spacetimes (i.e. for $\Lambda<0$) due to the differing boundary conditions used at spatial infinity. Thus in the following examples we will limit ourselves solely to $\Lambda\geq0$ cases. Nonetheless, the equations presented in section \ref{perturbations} remain valid for both positive and negative $\Lambda$. 

\subsection{Comparison to Other Results}

\subsubsection{Schwarzschild-de-Sitter}\label{sdssec}

We first consider the case of a unit mass Schwarzschild-de-Sitter black hole ($a=0$, $M=1$, $0\leq 9\Lambda \leq 1$) \cite{PhysRevD.15.2738}. The QNM frequencies for electromagnetic, gravitational, and massless scalar perturbations in a Schwarzschild-de-Sitter background have been calculated, for example, using the 6th order WKB method in \cite{Zhidenko:2003wq}. We find that in general the frequencies calculated using the analytic expansion method of \cite{2009CQGra..26v5003D} (utilising the expansion coefficients given in appendices \ref{scalarcoeff}-\ref{polarcoeff}) are in good agreement with those presented in \cite{Zhidenko:2003wq}. 

\begin{figure}
\caption{Complex massless scalar frequencies for the $n=0$ mode for varying values of $\Lambda$}
\label{scalarcomplanesds}
\includegraphics[width=0.5\textwidth]{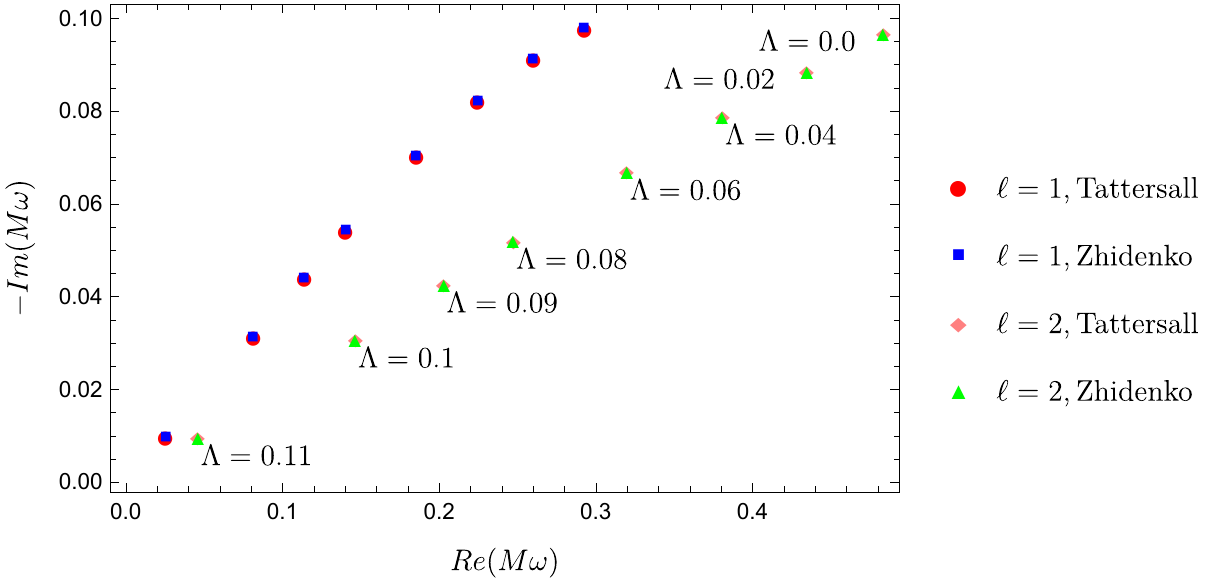}
\end{figure}

\begin{figure}
\caption{Complex electromagnetic frequencies for the $n=0$ mode for varying values of $\Lambda$}
\label{vectorcomplanesds}
\includegraphics[width=0.5\textwidth]{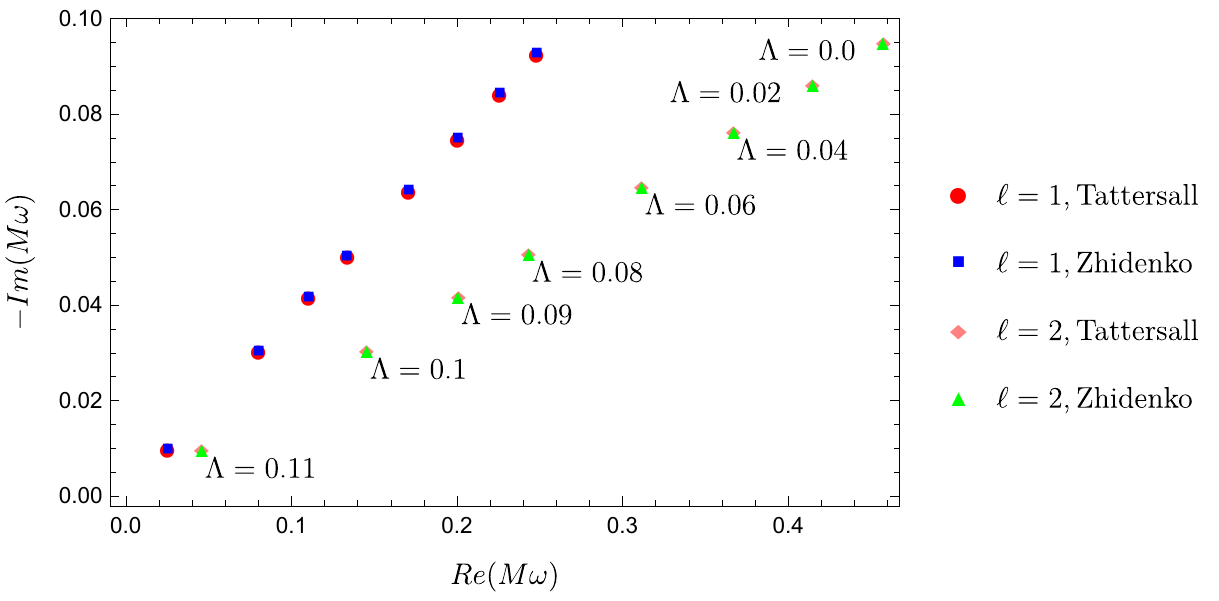}
\end{figure}

\begin{figure}
\caption{Complex gravitational frequencies for the $n=0$ mode for varying values of $\Lambda$}
\label{axialcomplanesds}
\includegraphics[width=0.5\textwidth]{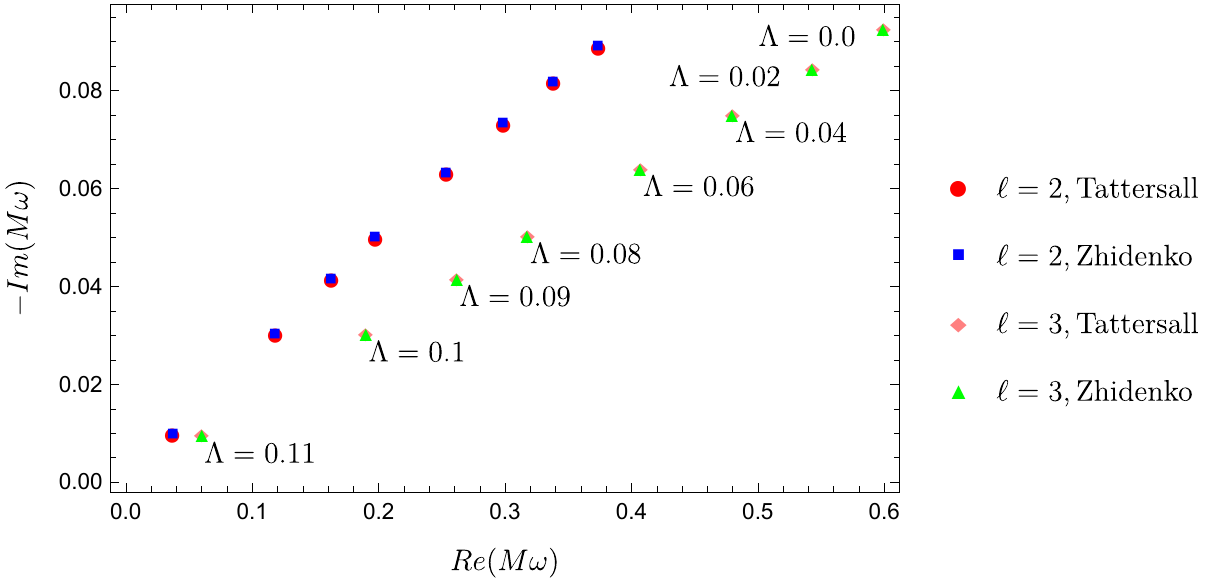}
\end{figure}

Tables \ref{sdstablescalar}--\ref{sdstableaxial} in appendix \ref{tables} give the QNM frequencies for massless scalar perturbations, massless vector (i.e. electromagnetic) perturbations, and gravitational perturbations for a selection of $\Lambda$ values. We do not distinguish between axial and polar gravitational frequencies in this case as, further to the discussion above, the coefficients presented in appendix \ref{polarcoeff} show that for $a=0$ the axial and polar QNM frequencies are isospectral (this was also shown in \cite{Zhidenko:2003wq}). In all cases the frequencies are presented as calculated via the WKB method as in \cite{Zhidenko:2003wq} and via the analytical expansion method of \cite{2009CQGra..26v5003D} used in this work, with the errors between the two methods also given. Figures \ref{scalarcomplanesds}-\ref{axialcomplanesds} show QNM frequencies in the complex plane as calculated via both methods. 

Tables \ref{sdstablescalar}--\ref{sdstableaxial} show that in all cases the relative errors between methods stays comfortably below 1\%, with figures \ref{scalarcomplanesds}-\ref{axialcomplanesds} showing almost complete alignment of frequencies in the complex plane. The expressions for the QNM frequencies given in appendix \ref{coeff} thus appear to compare very well with other methods of calculating QNM frequencies in the case that $a=0$. 


\subsubsection{Slowly rotating Kerr}

We will now consider the case of a unit mass slowly rotating Kerr black hole ($a>0$, $M=1$, $\Lambda=0$) \cite{Kerr:1963ud}. The QNM frequencies for electromagnetic, gravitational, and massless scalar perturbations in a Kerr background have been calculated, for example, in \cite{Berti:2009kk} using Leaver's continued fraction method \cite{1985RSPSA.402..285L}. We will compare the results of using the expansion coefficients given in appendix \ref{coeff} to the QNM frequency data provided at \url{http://www.phy.olemiss.edu/~berti/ringdown/}. 

Tables \ref{kerrtablescalar}--\ref{kerrtablegrav} in appendix \ref{tables} give QNM frequencies for massless scalar perturbations, massless vector (i.e. electromagnetic) perturbations, and gravitational perturbations for a selection of $a$ values. As discussed above, we do not distinguish between axial and polar gravitational frequencies as to linear order in $a$ the axial and polar spectra are isospectral for $\Lambda=0$ (see appendix \ref{polarcoeff}). In all cases the frequencies are presented as calculated via continued fractions in \cite{Berti:2009kk} and via the analytical expansion method of \cite{2009CQGra..26v5003D} used in this work, with the errors between the two methods also given. In \cite{Tattersall:2018nve} the analytic expansion results for a \textit{massive} scalar field on a slowly rotating Kerr background were compared to the continued fractions results of \cite{2006PhRvD..73l4040K}.

\begin{figure}
\caption{Real and imaginary components of the $\ell=m=2$, $n=0$ massless scalar mode for varying values of $a$}
\label{kerrscalarl1}
\includegraphics[width=0.5\textwidth]{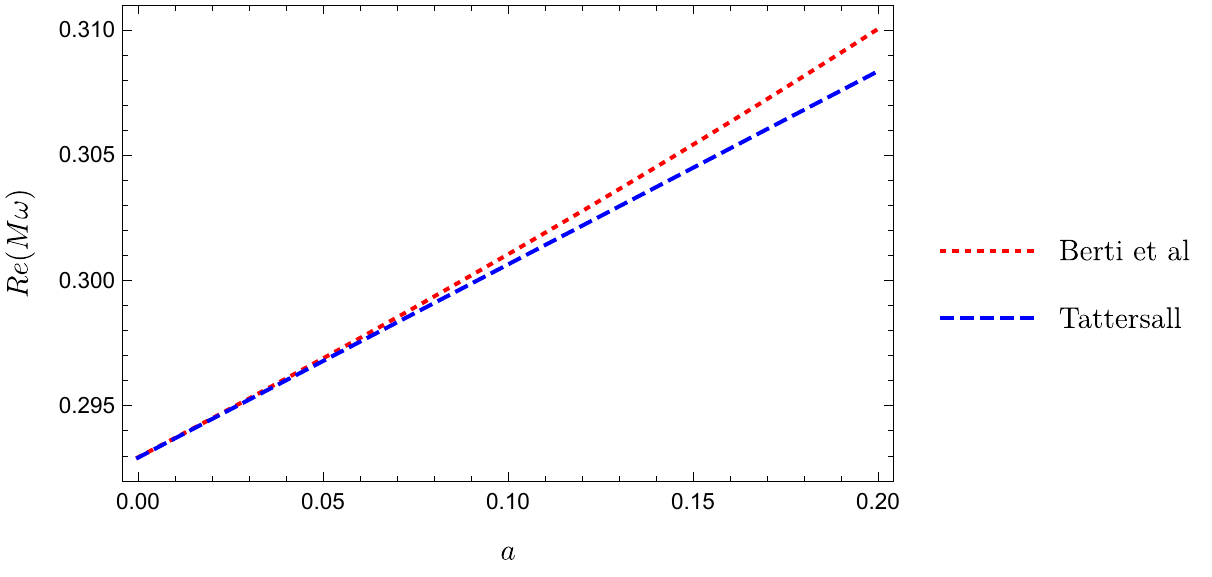}
\includegraphics[width=0.5\textwidth]{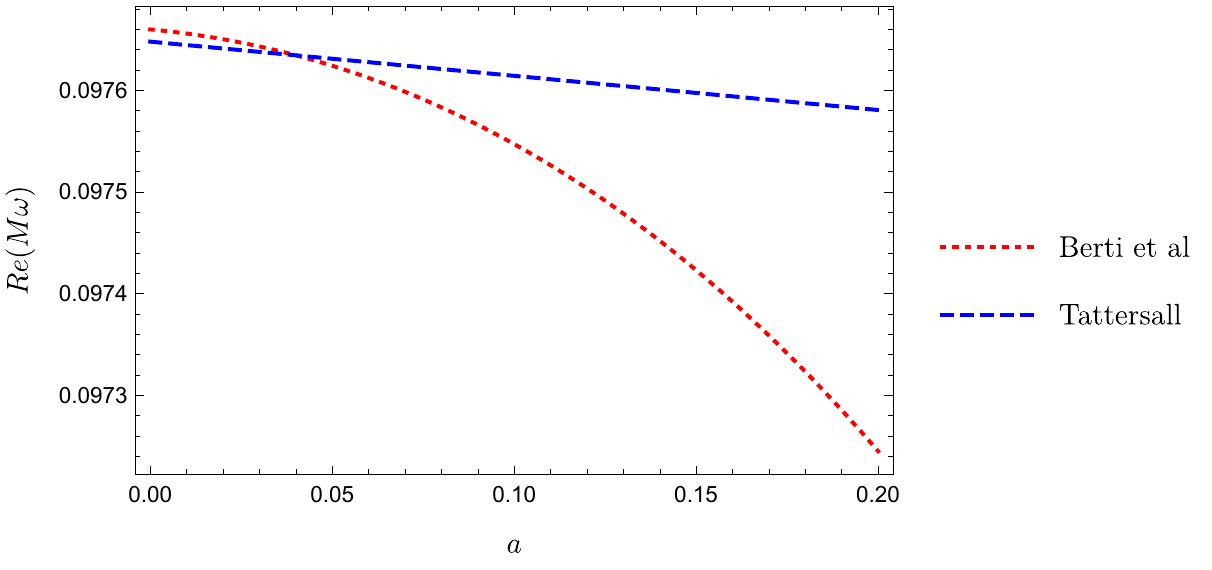}
\end{figure}

\begin{figure}
\caption{Real and imaginary components of the $\ell=m=2$, $n=0$ electromagnetic mode for varying values of $a$}
\label{kerrelecl2}
\includegraphics[width=0.5\textwidth]{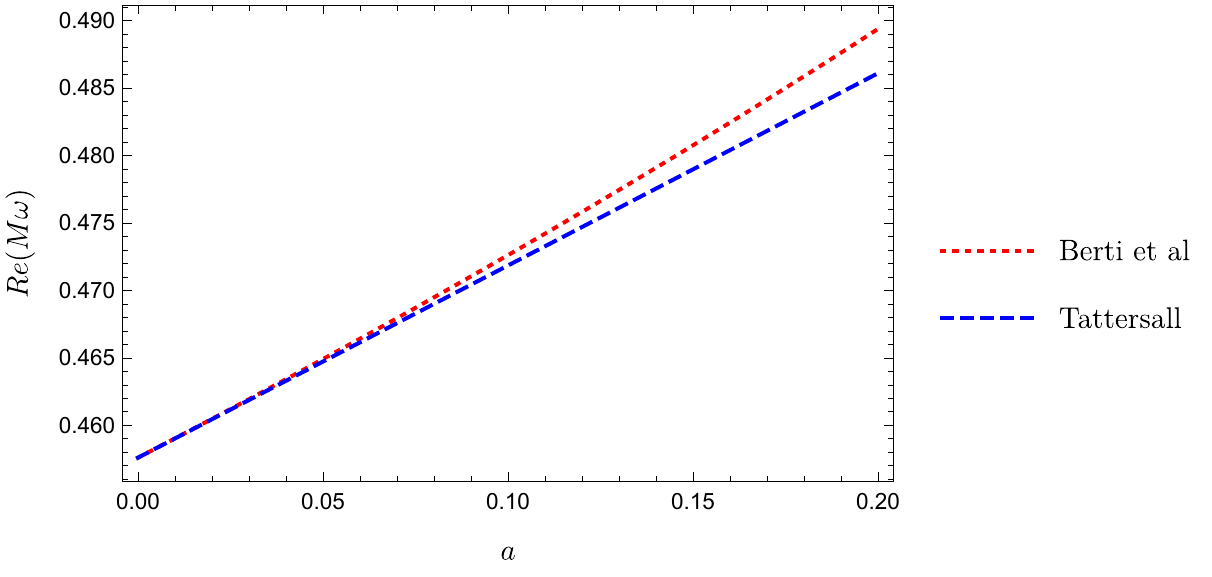}
\includegraphics[width=0.5\textwidth]{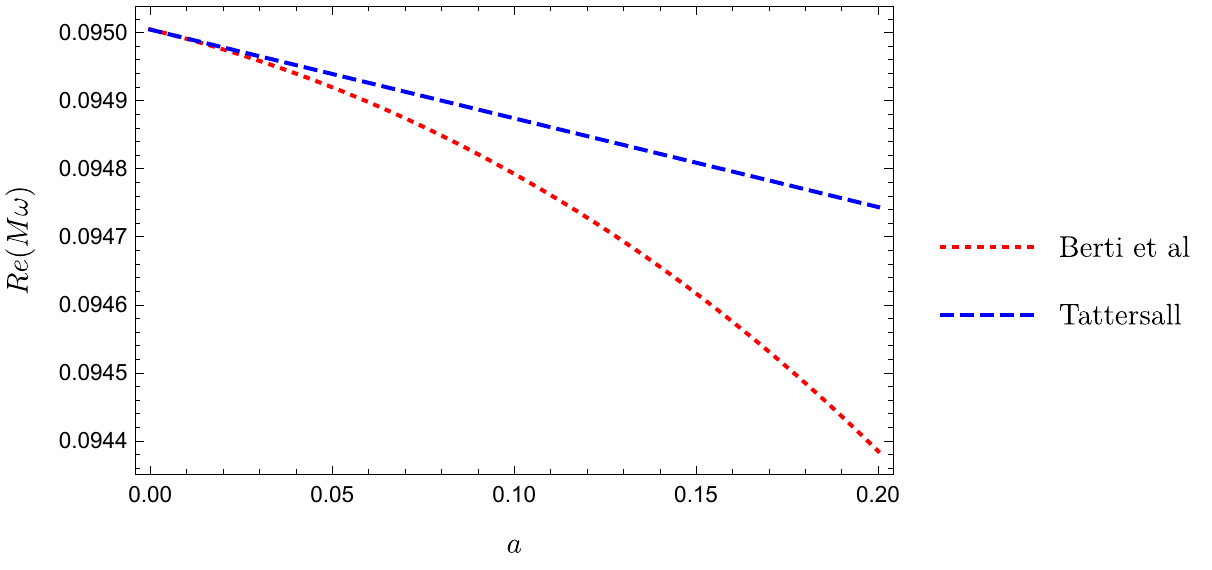}
\end{figure}

\begin{figure}
\caption{Real and imaginary components of the $\ell=m=3$, $n=0$ gravitational mode for varying values of $a$}
\label{kerrgravl3}
\includegraphics[width=0.5\textwidth]{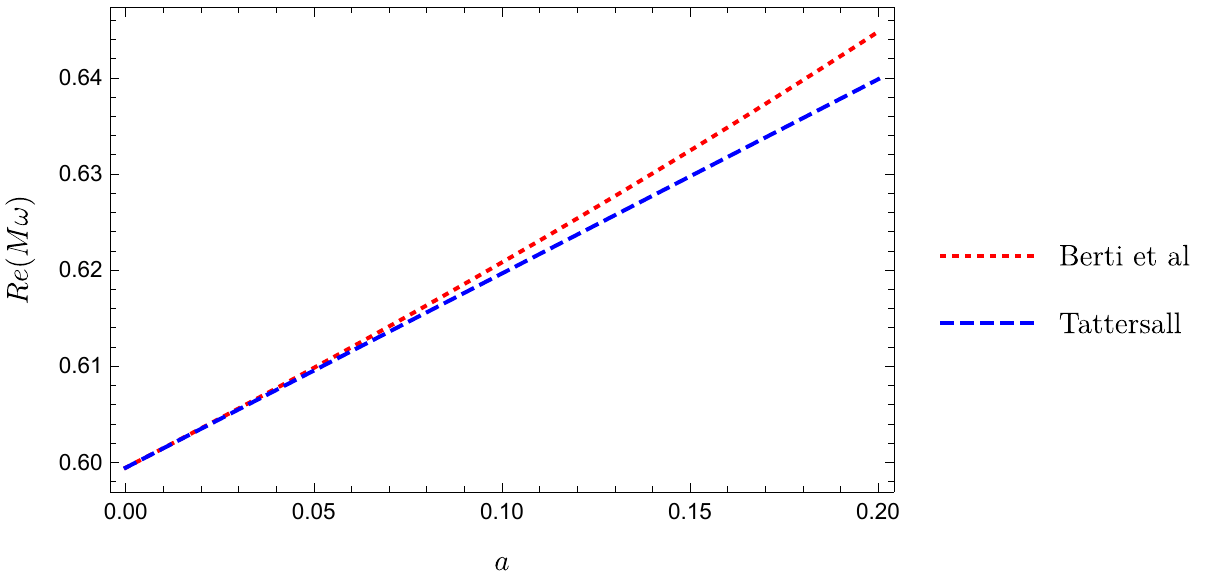}
\includegraphics[width=0.5\textwidth]{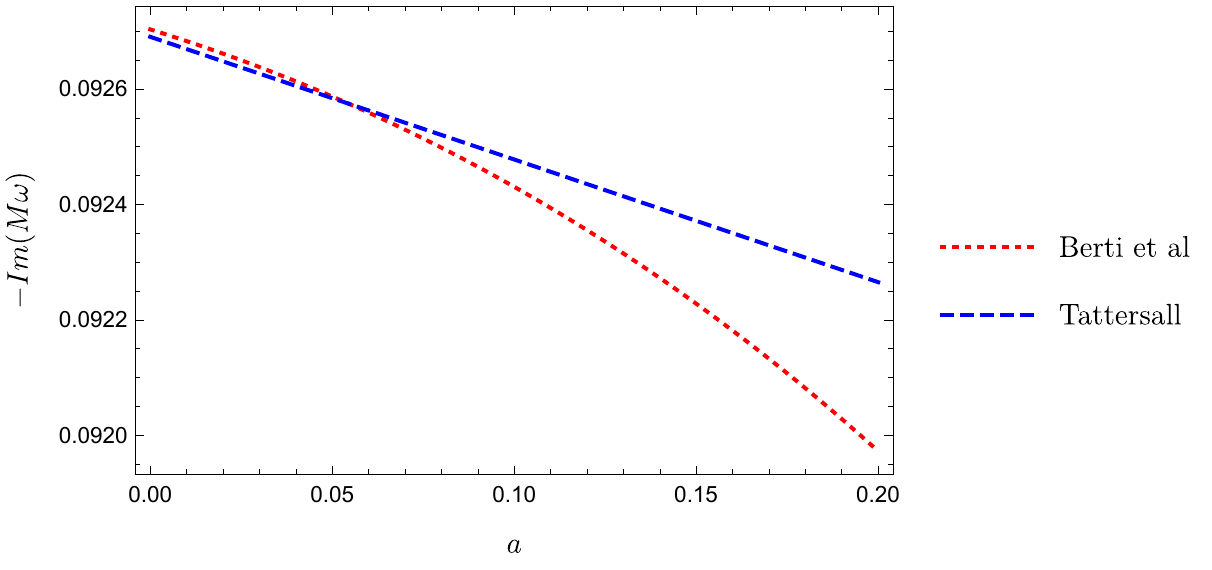}
\end{figure}

For $a\ll1$, figures \ref{kerrscalarl1}-\ref{kerrgravl3} show that the linear in $a$ approximation to the QNM frequencies calculated from the expressions in appendix \ref{coeff} match well with the numerical data, with tables \ref{kerrtablescalar}-\ref{kerrtablegrav} show the relative errors between methods staying below 1\% for spins of up to $a\sim 0.2$. From figures \ref{kerrscalarl1}-\ref{kerrgravl3}, however, we see that the linear in $a$ approximation for the QNM frequencies clearly starts to fail at smaller values of $a$. This is particularly noticeable for the imaginary frequency components, where the departure from the linear approximation is clearly seen by $a\sim 0.1$. Predictably, in all cases, as $a$ increases the difference between linear in spin approximation used in this work and the frequencies calculated numerically increases. 

The higher accuracy of the real frequency components compared to the imaginary components can be understood by considering that $O(a)$ contributions to the imaginary component only appear in two terms in the QNM expansion up to $O(L^{-6})$, $\omega_3$ and $\omega_5$ (see appendix \ref{coeff}). This is contrast to the real frequency component which receives contributions linear in $a$ in $\omega_0$, $\omega_2$, $\omega_4$, and $\omega_6$. Thus one should compute higher order terms in the expansion to calculate further $O(a)$ corrections to the imaginary component of the QNM frequencies, and thus improve agreement.

Nonetheless, for small $a$ the analytic expressions for the QNM frequencies appear to be a good approximation to those calculated numerically. 


\subsubsection{Slowly rotating Kerr-de Sitter}

We now turn to the most general case of a slowly rotating Kerr-dS black hole, where both $a$ and $\Lambda$ are non-zero and positive. The gravitational QNM frequencies for a Kerr-dS black hole of varying mass were calculated in \cite{PhysRevD.81.044005} by continued fractions, whilst their asymptotics were studied in \cite{Dyatlov:2011jd}.  Kerr-de Sitter black holes have further been studied in \cite{PhysRevD.97.104060}.

The expressions for $\omega_{-1}$ and $\omega_0$ in appendix \ref{coeff} (i.e. the first two terms in the $L$ expansion, as explained in section \ref{analyticsec}) agree exactly with the expressions given by eq.~(0.3) and (0.4) in \cite{Dyatlov:2011jd}, thus verifying the dominant $O(a)$ correction to $Re(\omega)$ calculated in this paper. 

Turning to comparisons with \cite{PhysRevD.81.044005}, a table of fitting parameters $(\bar{\omega}, \omega_a)$ (referred to as $(\omega_0, \omega_1)$ in \cite{PhysRevD.81.044005}) is provided to approximate the QNM frequencies of slowly rotating Kerr-dS black holes in the form:
\begin{align}
\omega = \bar{\omega} + am \omega_a +O(a^2),
\end{align}
where $\bar{\omega}$ is the QNM frequency evaluated for $a=0$, i.e. the corresponding Schwarzschild-dS frequency for a black hole of the same $M$ and $\Lambda$; $\omega_a$ represents the linear in $a$ correction term.

Table \ref{kerrdstable} compares data for $\bar{\omega}$ and $\omega_a$ from \cite{PhysRevD.81.044005} and as calculated in this paper for two black holes of differing mass, with the relative errors between both methods given in parentheses. Note that the fitting parameters $(\bar{\omega}, \omega_a)$ are rescaled by $M$ from their values in \cite{PhysRevD.81.044005} to coincide with the dimensionless black hole spin $a$ that we are using in this paper. Also note that, as previously mentioned, in the case that both $a$ and $\Lambda$ are non-zero, the axial and polar gravitational frequencies are no longer isospectral (see appendix \ref{polarcoeff}). Thus in table \ref{kerrdstable} we compare the results of \cite{PhysRevD.81.044005} to both axial and polar frequencies as calculated in this paper. 

\begin{table*}
\caption{Comparison of the $n=0$ gravitational QNM frequency fitting parameters as calculated by Leaver's continued fraction method in \cite{PhysRevD.81.044005} and analytical expansion techniques for varying $M$ with $\Lambda=3, \ell=2$.}
\label{kerrdstable}
\begin{ruledtabular}
\begin{tabular}{ccccccc}
\multicolumn{1}{l}{} &
\multicolumn{2}{c}{Leaver}&
\multicolumn{2}{c}{L-expansion (axial)}&
\multicolumn{2}{c}{L-expansion (polar)}\\
  $M$ & $\bar{\omega}$ & $\omega_a$  & $\bar{\omega}$ & $\omega_a$ & $\bar{\omega}$ & $\omega_a$
\\
\hline
\\
	 $0.1205$ &  $2.418-0.5944i$ & $0.67564+0.0092i$   & $2.419-0.5931i$ & $0.67391+0.0117i$  & $2.419-0.5931i$   & $0.67720+0.0112i$  \\
	(\% error)		&				&				& ($+0.04$,$-0.22$)& ($-0.26$,$+27$)&($+0.04$,$-0.22$)&($+0.23$,$+22$)\\
	$0.1789$ & $0.769-0.1964i$ & $0.58375+0.0022i$   & $0.770-0.1962i$ &  $0.58324+0.0024i$ &  $0.770-0.1962i$  & $0.58380+0.0022i$ \\
	(\% error)		&				&				& ($+0.13$,$-0.10$)& ($-0.09$,$+9.1$)& ($+0.13$,$-0.10$)&($+0.01$,$+0.00$)\\
\end{tabular}
\end{ruledtabular}
\end{table*}

As demonstrated in section \ref{sdssec} for the case of a Schwarzschild-dS black hole, we find good agreement between the $\bar{\omega}$ calculated in \cite{PhysRevD.81.044005} and those calculated in this work. We also find good agreement in the real part of $\omega_a$, with errors staying comfortably below 1\% when considering either the polar or axial frequencies. The imaginary part of $\omega_a$ does not agree as well, with relative errors of order $\sim10-20$\%. This can be understood similarly to the case of the imaginary Kerr frequencies as discussed above. For the imaginary frequency component, $O(a)$ contributions only appear in two terms in the QNM expansion up to $O(L^{-6})$ ($\omega_3$ and $\omega_5$). This is in contrast to the real frequency component which receives contributions linear in $a$ in all $\omega_{n}$ for even $n$. One should compute higher order terms in the expansion to calculate further $O(a)$ corrections to the imaginary component of the QNM frequencies. Indeed if one calculates, for example, $\omega_7$ for axial gravitational modes, the error in the imaginary component of $\omega_a$ for $M=0.1789$ drops from $9.1$\% to $5.1$\%, whilst for $M=0.1205$ the error drops from $27$\% to $11$\%. See appendix \ref{coeff} for the explicit expressions of the QNM frequencies calculated in this paper to $O(L^{-6})$.


\section{Conclusion}\label{conclusion}

In this paper we have presented Schr{\"o}dinger style master equations for the perturbations of massive scalar, massive vector, and gravitational fields on a slowly rotating Kerr-(A)dS black hole. These represent generalisations of the Regge-Wheeler and Zerilli equations (for fields of spin $0,-1,$ or $-2$) to include the effects of both a non-zero cosmological constant $\Lambda$ \textit{and} of slow rotation (i.e. to linear order in dimensionless black hole spin $a$). Some of these equations have been presented before in their entirety (e.g. eq.~(\ref{scalareq}) and (\ref{procaeq}) in \cite{2012PhRvD..86j4017P}), whilst versions of the equations with either $a=0$ or $\Lambda=0$ have been presented in, for example, \cite{Pani:2013pma,Cardoso:2001bb}. The generalisation of the gravitational perturbation equations to include both the effects of rotation and of a cosmological constant presented here should, however, prove useful given the wealth of knowledge accumulated to address such Schr{\"o}dinger style equations. 

We have also presented, following the method of \cite{2009CQGra..26v5003D}, analytical expressions for the QNM frequencies that satisfy each of the perturbation master equations present in section \ref{perturbations} (eq.~(\ref{scalareq}), (\ref{procaeq}), (\ref{gravaxial}), and (\ref{gravpolar})). The expressions given in appendix \ref{coeff} are intended as a compliment to existing methods of QNM calculation, with the equations presented in section \ref{perturbations} of course being amenable to being solved via one's preferred method. Given that there are relatively few numerical results for QNM frequencies in the literature for some categories of black holes (e.g. Kerr-de Sitter), numerically investigating the perturbation master equations presented in this paper and elsewhere is worthy of further attention. In addition, a natural extension of this work would be to consider black holes possessing non-zero electric charge \cite{doi:10.1063/1.1704350,doi:10.1063/1.1704351}. 

In section \ref{qnmsection} we find that the analytic expressions calculated in this paper agree well with the QNM frequencies calculated via other methods for a Schwarzschild-dS black hole \cite{Zhidenko:2003wq}, for a slowly rotating Kerr black hole \cite{Berti:2009kk}, and for a slowly rotating Kerr-de Sitter black hole \cite{PhysRevD.81.044005,Dyatlov:2011jd}. They are not, however, valid for asymptotically AdS spacetimes (as explained in \cite{2009CQGra..26v5003D}). The frequencies calculated in this paper support the numerically observed isospectrality of gravitational modes to linear order in spin for Kerr black holes \cite{Pani:2013wsa}, whilst the axial and polar gravitational spectra are shown to split for $a\neq0$ \textit{and} $\Lambda \neq0$. Given the good agreement with numerical results, the analytic expressions for QNM frequencies presented here provide a useful addition to those techniques already in the modern physicist's toolbox, allowing one to see the explicit dependence of the QNM frequencies on the parameters of the black hole and/or field. 

The study of gravitational QNM frequencies is, of course, of great interest in the context of gravitational wave observations. Properties of black hole merger remnants can be inferred from the observation of the QNM ringing, as well as tests of GR and of the `no-hair hypothesis' \cite{Dreyer:2003bv,Berti:2005ys,Berti:2015itd,0264-9381-33-17-174001,2018PhRvD..97d4021T,Berti:2018vdi,Barack:2018yly,Brito:2018rfr}. Meanwhile, the study of the QNM frequencies of massive bosons (e.g. the scalar and vector cases considered here) propagating on rotating black hole backgrounds finds great relevance in the study of black hole superradiance \cite{Brito:2015oca}. Furthermore, the AdS/CFT correspondence continues to provide motivation for studying QNMs in asymptotically AdS spacetimes \cite{Maldacena:1997re,Nunez:2003eq,Son:2007vk,Hartnoll:2009sz,Herzog:2009xv}.

The technique of recasting the complicated, multidimensional, equations of motion governing black hole perturbations in GR into decoupled one-dimensional Schr{\"o}dinger style equations is an incredibly useful one that has allowed great advances in the understanding and numerical calculation of QNM frequencies. The ability to execute such a simplification of the equations of motion, and in particular to include the effects of rotation, in theories of gravity \textit{beyond} GR is in many cases still a work in progress \cite{Tattersall:2018nve,Konoplya:2001ji,Cardoso:2009pk,Molina:2010fb,Lasky:2010bd,Brito:2013wya,Brito:2013yxa,Babichev:2015zub,Blazquez-Salcedo:2016enn,Blazquez-Salcedo:2017txk,Dong:2017toi,2018arXiv180709081B,Zhang:2014kna}. Given that the strong gravity regime of black hole mergers is likely to be one of the best `laboratories' available to us to probe any potential deviations from Einstein's theory, continuing the analysis of black hole perturbations for a variety of fields in alternative theories of gravity will remain an important avenue of research as gravitational wave astronomy matures in the coming years.  


\section*{Acknowledgments}
\vspace{-0.2in}
\noindent The author would like to thank P. G. Ferreira, A. O. Starinets, E. Berti, and S. Dyatlov for insightful discussions and advice throughout the preparation of this paper. OJT was supported by the Science and Technology Facilities Council (STFC) Project Reference 1804725, and acknowledges financial support from ERC Grant No: 693024.


\appendix

\section{Polar Gravitational potential}\label{appendixpolar}
The $O(a)$ correction to the potential for polar gravitational perturbations featured in eq.~(\ref{gravpolar}) is given by:
\begin{widetext}
\begin{align}
V^{(1)}_Z(r)=\;&\frac{F(r)^{-1}M}{729 r^8 \omega  \ell  (\ell +1) \left(6 M+r \left(\ell ^2+\ell -2\right)\right)^4} \left(486 l (l+1) \left(6 M+\left(l^2+l-2\right) r\right)^4 \left(\Lambda  r^3+6 M\right) \omega ^2 r^5+9 \Lambda  \left(\Lambda  r^3-3 r+6 M\right)\right.\nonumber\\
&\times
   \left(\left(l^2+l-2\right)^3 \left(5 r^2 \Lambda -3\right) \left(12 \omega ^2+\left(l^2+l+4\right) \Lambda \right) r^8+6 \left(l^2+l-2\right)^2\right.\nonumber\\
   &\times M \left(3
   l^4+6 l^3+\left(16 \Lambda ^2 r^4-6 \left(2 \omega ^2+5 \Lambda \right) r^2-45\right) l^2+2 \left(8 \Lambda ^2 r^4-3 \left(2 \omega ^2+5 \Lambda \right)
   r^2-24\right) l\right.\nonumber\\
   &\left.+4 \left(3 \Lambda ^3 r^6+\Lambda  \left(54 \omega ^2+11 \Lambda \right) r^4-6 \left(5 \omega ^2+2 \Lambda \right) r^2-6\right)\right)
   r^5\nonumber\\
   &-12 \left(l^2+l-2\right) M^2 \left(6 l^6+18 l^5-72 l^4-174 l^3+3 \left(-14 \Lambda ^2 r^4+9 \left(4 \omega ^2+5 \Lambda \right) r^2-123\right) l^2\right.\nonumber\\
   &\left.+3
   \left(-14 \Lambda ^2 r^4+9 \left(4 \omega ^2+5 \Lambda \right) r^2-93\right) l+4 r^2 \left(108 \omega ^2+\Lambda  \left(81-2 r^2 \left(99 \omega ^2+\Lambda
    \left(13 r^2 \Lambda -6\right)\right)\right)\right)+708\right) r^4\nonumber\\
    &+72 M^3 \left(6 l^6+18 l^5-381 l^4-792 l^3+\left(8 \left(4 \Lambda  \left(\Lambda 
   r^2+9\right)-9 \omega ^2\right) r^2+1179\right) l^2+2 \left(4 \left(4 \Lambda  \left(\Lambda  r^2+9\right)-9 \omega ^2\right) r^2+789\right) l\right.\nonumber\\
   &\left.+4 \left(20
   \Lambda ^3 r^6+\Lambda  \left(81 \omega ^2-73 \Lambda \right) r^4-9 \left(5 \omega ^2+13 \Lambda \right) r^2-402\right)\right) r^3\nonumber\\
   &+432 M^4 \left(80 \Lambda
   ^2 r^4-12 (8 l (l+1)-13) \Lambda  r^2+3 (l-1) (l+2) (37 l (l+1)-144)\right) r^2\nonumber\\
   &\left.+31104 M^5 \left(-2 \Lambda  r^2+8 l (l+1)-19\right) r+373248 M^6\right)
   r^3-2 \left(6 M+\left(l^2+l-2\right) r\right) \Lambda  \left(3 M-r^3 \Lambda \right) \left(\Lambda  r^3-3 r+6 M\right)^3\nonumber\\
   &\times \left(-\left(l^2+l-2\right)^2
   \left(12 \omega ^2+\left(l^2+l+4\right) \Lambda \right) r^6-3 \left(l^2+l-2\right) M \left(l^4+2 l^3+\left(4 r^2 \Lambda -15\right) l^2+4 \left(r^2 \Lambda
   -4\right) l\right.\right.\nonumber\\
   &\left.+4 \left(\Lambda ^2 r^4+3 \left(5 \omega ^2+\Lambda \right) r^2-2\right)\right) r^3+6 M^2 \left(3 l (l+1) \left(l^2+l-26\right)-4 \left(5
   \Lambda ^2 r^4-9 \left(\Lambda -3 \omega ^2\right) r^2-36\right)\right) r^2\nonumber\\
   &\left.+432 M^3 \left(3 \left(l^2+l-2\right)-2 r^2 \Lambda \right) r+2592 M^4\right)
   r-54 \left(\Lambda  r^3-3 r+6 M\right) \nonumber\\
   &\times\left(\left(l^2+l-2\right)^3 \Lambda  \left(-3 \left(l^2+l+2 r^2 \Lambda -6\right) \omega ^2-l (l+1) \Lambda 
   \left(l^2+l+r^2 \Lambda -5\right)\right) r^{11}\right.\nonumber\\
   &+\left(l^2+l-2\right)^2 M \left(3 \left(5 r^2 \Lambda -18\right) l^6+9 \left(5 r^2 \Lambda -18\right) l^5-3
   \left(7 \Lambda ^2 r^4+2 \left(\Lambda -6 \omega ^2\right) r^2+18\right) l^4\right.\nonumber\\
   &+\left(-42 \Lambda ^2 r^4+\left(72 \omega ^2-87 \Lambda \right) r^2+162\right)
   l^3-\left(r^2 \left(36 \left(2 \Lambda  r^2+5\right) \omega ^2+\Lambda  \left(\Lambda  \left(14 r^2 \Lambda -81\right) r^2+99\right)\right)-324\right)
   l^2\nonumber\\
   &+2 \left(\left(\Lambda  \left(r^2 \Lambda  \left(51-7 r^2 \Lambda \right)-24\right)-36 \left(\Lambda  r^2+3\right) \omega ^2\right) r^2+108\right) l\nonumber\\
   &\left.+4
   r^2 \left(9 \left(8-3 r^2 \Lambda  \left(r^2 \Lambda -4\right)\right) \omega ^2+2 \Lambda  \left(\Lambda  \left(r^2 \Lambda -3\right)
   r^2+9\right)\right)-432\right) r^6\nonumber\\
   &+6 \left(l^2+l-2\right) M^2 \left(21 l^8+84 l^7+3 \left(9 r^2 \Lambda -10\right) l^6+\left(81 r^2 \Lambda -384\right)
   l^5+\left(9 r^2 \left(-4 r^2 \Lambda ^2+\Lambda +16 \omega ^2\right)-303\right) l^4\right.\nonumber\\
   &+3 \left(44-3 r^2 \left(\Lambda  \left(8 \Lambda  r^2+13\right)-32
   \omega ^2\right)\right) l^3-6 \left(r^2 \left(4 \left(8 \Lambda  r^2+21\right) \omega ^2+\Lambda  \left(r^2 \Lambda -10\right) \left(r^2 \Lambda
   -1\right)\right)-128\right) l^2\nonumber\\
   &-6 \left(r^2 \left(4 \left(8 \Lambda  r^2+27\right) \omega ^2+\Lambda  \left(r^2 \Lambda  \left(r^2 \Lambda
   -17\right)-2\right)\right)-104\right) l\nonumber\\
   &\left.+4 \left(r^2 \left(3 \left(\Lambda  \left(59-8 r^2 \Lambda \right) r^2+60\right) \omega ^2+\Lambda  \left(\Lambda ^3
   r^6-15 \Lambda  r^2+12\right)\right)-228\right)\right) r^5\nonumber\\
   &+12 M^3 \left(90 l^8+360 l^7+\left(75 r^2 \Lambda -36\right) l^6+9 \left(25 r^2 \Lambda
   -152\right) l^5-3 \left(21 \Lambda ^2 r^4+8 \left(5 \Lambda -21 \omega ^2\right) r^2+162\right) l^4\right.\nonumber\\
   &+3 \left(\left(336 \omega ^2-\Lambda  \left(42 \Lambda 
   r^2+205\right)\right) r^2+576\right) l^3+3 \left(\left(\Lambda  \left(\Lambda  \left(4 \Lambda  r^2+57\right) r^2+27\right)-12 \left(17 \Lambda 
   r^2+42\right) \omega ^2\right) r^2+192\right) l^2\nonumber\\
   &+6 \left(r^2 \left(\Lambda  \left(\Lambda  \left(2 \Lambda  r^2+39\right) r^2+71\right)-6 \left(17 \Lambda
    r^2+56\right) \omega ^2\right)-96\right) l\nonumber\\
    &\left.+4 \left(r^2 \left(9 \left(\Lambda  \left(43-2 r^2 \Lambda \right) r^2+56\right) \omega ^2+\Lambda  \left(r^2
   \Lambda  \left(5 r^2 \Lambda  \left(r^2 \Lambda -6\right)-27\right)-18\right)\right)-72\right)\right) r^4\nonumber\\
   &+72 M^4 \left(87 l^6+261 l^5+9 \left(7 r^2 \Lambda
   -57\right) l^4+3 \left(42 r^2 \Lambda -487\right) l^3+3 \left(\left(72 \omega ^2-\Lambda  \left(12 \Lambda  r^2+25\right)\right) r^2+516\right) l^2\right.\nonumber\\
   &\left.-6
   \left(r^2 \left(\Lambda  \left(6 \Lambda  r^2+23\right)-36 \omega ^2\right)-387\right) l+2 r^2 \left(2 \Lambda  \left(\Lambda  \left(11 r^2 \Lambda
   -9\right) r^2+6\right)-9 \left(11 \Lambda  r^2+24\right) \omega ^2\right)-2244\right) r^3\nonumber\\
   &+2592 M^5 \left(21 l^4+42 l^3+\left(2 r^2 \Lambda -127\right)
   l^2+2 \left(r^2 \Lambda -74\right) l+2 \left(\Lambda ^2 r^4+\left(3 \omega ^2+\Lambda \right) r^2+115\right)\right) r^2\nonumber\\
   &\left.\left.+5184 M^6 \left(-2 \Lambda  r^2+54 l
   (l+1)-171\right) r+497664 M^7\right)\right)\end{align}
\end{widetext}

\section{Quasinormal frequency expansion coefficients}\label{coeff}

As explained in Section \ref{qnmsection} and in \cite{2009CQGra..26v5003D}, the fundamental (i.e. overtone index $n=0$) quasinormal frequencies $\omega$ satisfying eq.~(\ref{scalareq}) (for scalar perturbations),  eq.~(\ref{procaeq}) (for axial vector perturbations), eq.~(\ref{gravaxial}) (for axial gravitational perturbations), and for eq.~(\ref{gravpolar}) (for polar gravitational perturbations) can be expressed as a power series in inverse powers of $L=\ell+1/2$:
\begin{align}
\omega=\sum_{k=-1}^{k=\infty}\omega_k L^{-k}.
\end{align}
The expansion coefficients $\omega_k$ for each of the perturbation types (scalar, vector, or tensor), to $O(L^{-6})$, are given below. Note that, as explained in \cite{2009CQGra..26v5003D}, the below expressions are valid only for $\Lambda\geq0$.

\subsection{Scalar frequencies}\label{scalarcoeff}

The $\omega_k$ that satisfy eq.~(\ref{scalareq}) are given by:
\begin{widetext}
\begin{align}
3\sqrt{3}M\omega_{-1}=&\;\sqrt{1-9\Lambda M^2}\nonumber\\
3\sqrt{3}M\omega_{0}=&\;-\frac{i}{2}\sqrt{1-9\Lambda M^2}+am\sqrt{3}\left(\frac{2}{9}+\Lambda M^2\right)\nonumber\\
3\sqrt{3}M\omega_{1}=&\;\frac{81 \Lambda  M^4 \left(61 \Lambda -108 \mu ^2\right)+M^2 \left(972 \mu ^2-612 \Lambda \right)+7}{216 \sqrt{1-9 \Lambda  M^2}}\nonumber\\
3\sqrt{3}M\omega_{2}=&\;-\frac{i \left(1-9 \Lambda  M^2\right)^{3/2} \left(45 M^2 \left(401 \Lambda -648 \mu ^2\right)+137\right)}{7776}-a m\frac{ \left(81 \Lambda  M^4 \left(61 \Lambda -108 \mu ^2\right)+M^2 \left(972 \mu ^2-450 \Lambda \right)-11\right)}{162 \sqrt{3}}\nonumber\\
3\sqrt{3}M\omega_{3}=&\;\frac{1}{2519424}\left(\sqrt{1-9 \Lambda  M^2} \left(729 \Lambda  M^6 \left(750851 \Lambda ^2-905904 \Lambda  \mu ^2-419904 \mu ^4\right)\right.\right.\nonumber\\
&\left.\left.-1944 M^4 \left(53119 \Lambda ^2-81567 \Lambda  \mu ^2-4374 \mu
   ^4\right)+27 M^2 \left(146681 \Lambda -245592 \mu ^2\right)+5230\right)\right)\nonumber\\
   &-i a m\frac{ \left(1-9 \Lambda  M^2\right) \left(405 \Lambda  M^4 \left(401 \Lambda -648 \mu ^2\right)-36 M^2 \left(332 \Lambda -567 \mu ^2\right)-29\right)}{2916 \sqrt{3}}\nonumber\\
3\sqrt{3}M\omega_{4}=&\;\frac{i}{362797056}\left( \left(1-9 \Lambda  M^2\right)^{3/2} \left(3645 \Lambda  M^6 \left(27099013 \Lambda ^2-7301664 \Lambda  \mu ^2-52907904 \mu ^4\right)\right.\right.\nonumber\\
&\left.\left.-243 M^4 \left(68373857 \Lambda
   ^2-101570112 \Lambda  \mu ^2-7558272 \mu ^4\right)+27 M^2 \left(12794177 \Lambda -20816352 \mu ^2\right)+590983\right)\right)\nonumber\\
   &-a m\frac{1}{629856 \sqrt{3}} \left(\left(1-9 \Lambda  M^2\right) \left(27 M^2 \left(-11337408 \Lambda  \mu ^4 M^4-3888 \mu ^2 \left(3 \Lambda  M^2 \left(2097 \Lambda  M^2-437\right)+35\right)\right.\right.\right.\nonumber\\
   &\left.\left.\left.+\Lambda  \left(9
   \Lambda  M^2 \left(2252553 \Lambda  M^2-357239\right)+80915\right)\right)+8137\right)\right)\nonumber\\
 3\sqrt{3}M\omega_{5}=&\;  \frac{\sqrt{1-9 \Lambda  M^2}}{39182082048} \left(9 M^2 \left(22674816 \mu ^4 M^2 \left(27 \Lambda  M^2 \left(144 \Lambda  M^2 \left(6501 \Lambda  M^2-1363\right)+9907\right)-427\right)\right.\right.\nonumber\\
 &-2592 \mu ^2
   \left(9 \Lambda  M^2 \left(27 \Lambda  M^2 \left(3 \Lambda  M^2 \left(112055580 \Lambda  M^2-4940639\right)-7820726\right)+16693957\right)-1450348\right)\nonumber\\
   &+\Lambda  \left(-9 \Lambda 
   M^2 \left(9 \Lambda  M^2 \left(9 \Lambda  M^2 \left(168232451787 \Lambda  M^2-63977771143\right)+71553627542\right)-28276533542\right)\right.\nonumber\\
   &\left.\left.\left.-2271718855\right)+7346640384 \mu ^6 M^4
   \left(36 \Lambda  M^2 \left(198 \Lambda  M^2-17\right)+7\right)\right)-42573661\right)\nonumber\\
   &+i a m\frac{1 \left(1-9 \Lambda  M^2\right)}{68024448 \sqrt{3}} \left(9 M^2 \left(-34012224 \mu ^4 M^2 \left(9 \Lambda  M^2 \left(630 \Lambda  M^2-59\right)-2\right)\right.\right.\nonumber\\
   &-17496 \mu ^2 \left(3 \Lambda  M^2
   \left(9 \Lambda  M^2 \left(56340 \Lambda  M^2-55807\right)+48878\right)-1721\right)\nonumber\\
   &\left.\left.+\Lambda  \left(27 \Lambda  M^2 \left(3658366755 \Lambda ^2 M^4-911149638 \Lambda 
   M^2+59613232\right)-18218222\right)\right)+41735\right)\nonumber\\
  3\sqrt{3}M\omega_{6}=&\; \frac{i \sqrt{1-9 \Lambda  M^2}}{8463329722368} \left(27 M^2 \left(136048896 \mu ^4 M^2 \left(1-9 \Lambda  M^2\right) \left(243 \Lambda  M^2 \left(15 \Lambda  M^2 \left(217247 \Lambda 
   M^2-41079\right)+22487\right)\right.\right.\right.\nonumber\\
   &\left.+6563\right)+\Lambda  \left(9 \Lambda  M^2 \left(3 \Lambda  M^2 \left(27 \Lambda  M^2 \left(315 \Lambda  M^2 \left(429275206029 \Lambda 
   M^2-215443481162\right)+11851932821509\right)\right.\right.\right.\nonumber\\
   &\left.\left.\left.-23397470018140\right)+1898828714953\right)-84181473166\right)\nonumber\\
   &-1296 \mu ^2 \left(1-9 \Lambda  M^2\right) \left(9 \Lambda  M^2 \left(135
   \Lambda  M^2 \left(43894090161 \Lambda ^2 M^4-4927747056 \Lambda  M^2-156842798\right)+2214937208\right)\right.\nonumber\\
   &\left.\left.\left.-100404965\right)-264479053824 \mu ^6 M^4 \left(9 \Lambda  M^2-1\right)
   \left(405 \Lambda  M^2 \left(313 \Lambda  M^2-18\right)+37\right)\right)+11084613257\right)\nonumber\\
   &+a m\frac{1}{29386561536 \sqrt{3}}  \left(27 M^2 \left(120932352 \mu ^4 M^2 \left(9 \Lambda  M^2-1\right) \left(27 \Lambda  M^2 \left(1485 \Lambda  M^2 \left(197 \Lambda 
   M^2-36\right)+2047\right)\right.\right.\right.\nonumber\\
   &\left.+277\right)-1728 \mu ^2 \left(9 \Lambda  M^2-1\right) \left(9 \Lambda  M^2 \left(27 \Lambda  M^2 \left(3 \Lambda  M^2 \left(280138950 \Lambda 
   M^2-5781821\right)-18236525\right)+29333473\right)\right.\nonumber\\
   &\left.-1464535\right)+\Lambda  \left(1551785558-9 \Lambda  M^2 \left(3 \Lambda  M^2 \left(27 \Lambda  M^2 \left(45 \Lambda  M^2
   \left(56077483929 \Lambda  M^2-25842254398\right)\right.\right.\right.\right.\nonumber\\
   &\left.\left.\left.\left.+194725826099\right)-383179273148\right)+32615632811\right)\right)\nonumber\\
   &\left.\left.-19591041024 \mu ^6 M^4 \left(1-9 \Lambda  M^2\right) \left(9
   \Lambda  M^2 \left(495 \Lambda  M^2-29\right)+1\right)\right)+18404153\right) 
\end{align}
\end{widetext}

\subsection{Axial Vector frequencies}\label{vectorcoeff}

The $\omega_k$ that satisfy eq.~(\ref{procaeq}) are given by:
\begin{widetext}
\begin{align}
3\sqrt{3}M\omega_{-1}=&\;\sqrt{1-9\Lambda M^2}\nonumber\\
3\sqrt{3}M\omega_{0}=&\;-\frac{i}{2}\sqrt{1-9\Lambda M^2}+am\sqrt{3}\left(\frac{2}{9}+\Lambda M^2\right)\nonumber\\
3\sqrt{3}M\omega_{1}=&\;\frac{1}{216} \sqrt{1-9 \Lambda  M^2} \left(M^2 \left(99 \Lambda +972 \mu ^2\right)-65\right)\nonumber\\
3\sqrt{3}M\omega_{2}=&\;\frac{5 i \left(1-9 \Lambda  M^2\right)^{3/2} \left(9 M^2 \left(31 \Lambda +648 \mu ^2\right)+59\right)}{7776}-am\frac{ \left(1-9 \Lambda  M^2\right)}{162 \sqrt{3}} \left(M^2 \left(99 \Lambda +972 \mu ^2\right)+25\right)\nonumber\\
3\sqrt{3}M\omega_{3}=&\;\frac{\sqrt{1-9 \Lambda  M^2}}{2519424} \left(27 M^2 \left(-314928 \mu ^4 M^2 \left(36 \Lambda  M^2-1\right)-648 \mu ^2 \left(9 \Lambda  M^2 \left(6786 \Lambda  M^2-935\right)+19\right)\right.\right.\nonumber\\
&\left.\left.+\Lambda\left(9 \Lambda  M^2 \left(6616-119127 \Lambda  M^2\right)+4121\right)\right)-71234\right)\nonumber\\
&+i a m\frac{\left(1-9 \Lambda  M^2\right)}{2916 \sqrt{3}} \left(9 M^2 \left(324 \mu ^2 \left(90 \Lambda  M^2-7\right)+\Lambda  \left(1395 \Lambda  M^2-76\right)\right)+245\right)\nonumber\\
3\sqrt{3}M\omega_{4}=&\;-\frac{i \left(1-9 \Lambda  M^2\right)^{3/2}}{362797056} \left(27 M^2 \left(68024448 \mu ^4 M^2 \left(105 \Lambda  M^2-1\right)+38880 \mu ^2 \left(27 \Lambda  M^2 \left(10011 \Lambda 
   M^2-698\right)-415\right)\right.\right.\nonumber\\
   &\left.\left.+\Lambda  \left(9 \Lambda  M^2 \left(19250805 \Lambda  M^2-288223\right)-585857\right)\right)-3374791\right)\nonumber\\
   &+am\frac{ \left(1-9 \Lambda  M^2\right)}{629856 \sqrt{3}} \left(27 M^2 \left(11337408 \Lambda  \mu ^4 M^4+3888 \mu ^2 \left(3 \Lambda  M^2 \left(3393 \Lambda  M^2-401\right)+11\right)\right.\right.\nonumber\\
   &\left.\left.+\Lambda  \left(9\Lambda  M^2 \left(119127 \Lambda  M^2-10825\right)-7043\right)\right)-21097\right)\nonumber\\
3\sqrt{3}M\omega_{5}=&\;\frac{\sqrt{1-9 \Lambda  M^2}}{39182082048} \left(9 M^2 \left(22674816 \mu ^4 M^2 \left(27 \Lambda  M^2 \left(144 \Lambda  M^2 \left(7689 \Lambda  M^2-1417\right)+8539\right)-67\right)\right.\right.\nonumber\\
&+2592 \mu ^2
   \left(9 \Lambda  M^2 \left(27 \Lambda  M^2 \left(3 \Lambda  M^2 \left(329310180 \Lambda  M^2-74177569\right)+13293734\right)-1411525\right)-173540\right)\nonumber\\
   &+\Lambda  \left(9 \Lambda 
   M^2 \left(63 \Lambda  M^2 \left(9 \Lambda  M^2 \left(1261925091 \Lambda  M^2-229959743\right)+55808902\right)+113404646\right)+160933625\right)\nonumber\\
   &\left.\left.+7346640384 \mu ^6 M^4 \left(36\Lambda  M^2 \left(198 \Lambda  M^2-17\right)+7\right)\right)-342889693\right)\nonumber\\
   &+i a m\frac{1}{68024448 \sqrt{3}} \left(9 M^2 \left(34012224 \mu ^4 M^2 \left(9 \Lambda  M^2-1\right) \left(9 \Lambda  M^2 \left(630 \Lambda  M^2-59\right)-2\right)\right.\right.\nonumber\\
   &+17496 \left(3 \Lambda  M^2 \left(600660\Lambda  M^2-30323\right)-871\right) \left(1 -9 \Lambda   M^2\right)^2\mu^2\nonumber\\
   &\left.\left.+\Lambda  \left(9 \Lambda  M^2 \left(9 \Lambda  M^2 \left(63 \Lambda  M^2 \left(8250345 \Lambda 
   M^2-2036507\right)+8064646\right)+424262\right)-3884077\right)\right)+3233783\right)\nonumber\\
3\sqrt{3}M\omega_{6}=&\;\frac{i \left(1-9 \Lambda  M^2\right)^{3/2}}{8463329722368} \left(9 M^2 \left(408146688 \mu ^4 M^2 \left(243 \Lambda  M^2 \left(15 \Lambda  M^2 \left(352463 \Lambda 
   M^2-44967\right)+13127\right)+6995\right)\right.\right.\nonumber\\
   &+3888 \mu ^2 \left(9 \Lambda  M^2 \left(27 \Lambda  M^2 \left(15 \Lambda  M^2 \left(25239221253 \Lambda 
   M^2-3989698864\right)+1860754534\right)+486103048\right)\right.\nonumber\\
   &\left.+90125093\right)+\Lambda  \left(9 \Lambda  M^2 \left(9 \Lambda  M^2 \left(45 \Lambda  M^2 \left(645007240719 \Lambda 
   M^2-78406459927\right)-380899298\right)+12521227898\right)\right.\nonumber\\
   &\left.\left.\left.+11148937343\right)+793437161472 \mu ^6 M^4 \left(405 \Lambda  M^2 \left(313 \Lambda 
   M^2-18\right)+37\right)\right)+74076561065\right)\nonumber\\
   &+am\frac{1}{29386561536 \sqrt{3}} \left(27 M^2 \left(120932352 \mu ^4 M^2 \left(9 \Lambda  M^2-1\right) \left(27 \Lambda  M^2 \left(135 \Lambda  M^2 \left(2563 \Lambda 
   M^2-408\right)+1777\right)\right.\right.\right.\nonumber\\
   &\left.+133\right)+1728 \mu ^2 \left(9 \Lambda  M^2-1\right) \left(9 \Lambda  M^2 \left(27 \Lambda  M^2 \left(3 \Lambda  M^2 \left(823275450 \Lambda 
   M^2-168911203\right)+27535325\right)-4529329\right)\right.\nonumber\\
   &\left.-1034153\right)+\Lambda  \left(9 \Lambda  M^2 \left(3 \Lambda  M^2 \left(27 \Lambda  M^2 \left(63 \Lambda  M^2 \left(2103208485
   \Lambda  M^2-613717798\right)+3480361741\right)\right.\right.\right.\nonumber\\
   &\left.\left.\left.\left.\left.-1868367940\right)-82065323\right)-87749482\right)+19591041024 \mu ^6 M^4 \left(9 \Lambda  M^2-1\right) \left(9 \Lambda  M^2
   \left(495 \Lambda  M^2-29\right)+1\right)\right)\right.\nonumber\\
   &\left.-82685575\right)\nonumber\\
\end{align}
\end{widetext}

\subsection{Axial Gravitational frequencies}\label{axialcoeff}

The $\omega_k$ that satisfy eq.~(\ref{gravaxial}) are given by:
\begin{widetext}
\begin{align}
3\sqrt{3}M\omega_{-1}=&\;\sqrt{1-9\Lambda M^2}\nonumber\\
3\sqrt{3}M\omega_{0}=&\;-\frac{i}{2}\sqrt{1-9\Lambda M^2}+am\sqrt{3}\left(\frac{2}{9}+\Lambda M^2\right)\nonumber\\
3\sqrt{3}M\omega_{1}=&\;\frac{1}{216} \sqrt{1-9 \Lambda  M^2} \left(99 \Lambda  M^2-281\right)\nonumber\\
3\sqrt{3}M\omega_{2}=&\;\frac{i \left(1-9 \Lambda  M^2\right)^{3/2} \left(1395 \Lambda  M^2+1591\right)}{7776}-am\frac{\left(1-9 \Lambda  M^2\right) \left(99 \Lambda  M^2+133\right)}{162 \sqrt{3}}\nonumber\\
3\sqrt{3}M\omega_{3}=&\;-\frac{\sqrt{1-9 \Lambda  M^2} \left(27 \Lambda  M^2 \left(9 \Lambda  M^2 \left(119127
   \Lambda  M^2+50408\right)+118999\right)+1420370\right)}{2519424}\nonumber\\
&+i a m\frac{\left(1-9 \Lambda  M^2\right) \left(279 \Lambda  M^2 \left(45 \Lambda 
   M^2+8\right)+893\right)}{2916 \sqrt{3}}\nonumber\\
3\sqrt{3}M\omega_{4}=&\;-\frac{i \left(1-9 \Lambda  M^2\right)^{3/2} \left(4677945615 \Lambda ^3 M^6+1687260051
   \Lambda ^2 M^4+211769829 \Lambda  M^2-92347783\right)}{362797056}\nonumber\\
&+am\frac{\left(1-9 \Lambda  M^2\right) \left(27 \Lambda  M^2 \left(9 \Lambda  M^2
   \left(119127 \Lambda  M^2+17687\right)-57587\right)+499895\right)}{629856 \sqrt{3}}\nonumber\\
3\sqrt{3}M\omega_{5}=&\;\frac{\sqrt{1-9 \Lambda  M^2}}{39182082048} \left(9 \Lambda  M^2 \left(9 \Lambda  M^2 \left(9 \Lambda 
   M^2 \left(63 \Lambda  M^2 \left(1261925091 \Lambda 
   M^2+181743169\right)-1078685462\right)-6276258970\right)\right.\right.\nonumber\\
   &\left.\left.-25334574535\right)-7827932509
   \right)\nonumber\\
   &-i a m\frac{\left(1-9 \Lambda  M^2\right) \left(9 \Lambda  M^2 \left(27 \Lambda  M^2 \left(3
   \Lambda  M^2 \left(57752415 \Lambda 
   M^2+5720786\right)-4178128\right)-42220738\right)+27500857\right)}{68024448 \sqrt{3}}\nonumber\\
3\sqrt{3}M\omega_{6}=&\;\frac{i \left(1-9 \Lambda  M^2\right)^{3/2}}{8463329722368}    \left(9 \Lambda  M^2 \left(9 \Lambda  M^2
   \left(9 \Lambda  M^2 \left(45 \Lambda  M^2 \left(645007240719 \Lambda 
   M^2+111206269001\right)-40406459618\right)\right.\right.\right.\nonumber\\
   &\left.\left.\left.-1321332614854\right)-1848252537217\right)-4
   81407154423\right)\nonumber\\
   &-am\frac{\left(1-9 \Lambda  M^2\right)}{29386561536 \sqrt{3}} \left(9 \Lambda  M^2 \left(9 \Lambda  M^2 \left(9
   \Lambda  M^2 \left(63 \Lambda  M^2 \left(6309625455 \Lambda 
   M^2+300876293\right)-8618549878\right)\right.\right.\right.\nonumber\\
   &\left.\left.\left.-12418517690\right)+11910847045\right)-615582833
   21\right)\label{axialcoeffeq}
\end{align}
\end{widetext}

\subsection{Polar Gravitational frequencies}\label{polarcoeff}

For the $\omega_k$ that satisfy eq.~(\ref{gravpolar}), we present the frequency coefficients in the form:
\begin{align}
\omega_k^{p}=\omega_k^{ax}+\Delta\omega_k,
\end{align}
where the $\omega_k^{ax}$ are given in eq.~(\ref{axialcoeffeq}), in order to clearly show under which circumstances isospectrality between the axial and polar gravitational sectors breaks down. The $\Delta\omega_k$ are given by:
\begin{widetext}
\begin{align}
\Delta\omega_{-1}=&\;0\nonumber\\
\Delta\omega_{0}=&\;0\nonumber\\
\Delta\omega_{1}=&\;0\nonumber\\
\Delta\omega_{2}=&\;0\nonumber\\
\Delta\omega_{3}=&\;0\nonumber\\
\Delta\omega_{4}=&\;\frac{am \Lambda M}{243}  \left(1-9 \Lambda  M^2\right) \left(9 \Lambda  M^2 \left(9 \Lambda  M^2 \left(27 \Lambda 
   M^2-11\right)-95\right)+148\right)\nonumber\\
\Delta\omega_{5}=&\;-\frac{i am \Lambda  M}{1458} \left(1-9 \Lambda  M^2\right) \left(81 \Lambda  M^2 \left(3 \Lambda  M^2 \left(3 \Lambda  M^2 \left(1188
   \Lambda  M^2-487\right)-296\right)+182\right)-152\right)\nonumber\\
\Delta\omega_{6}=&\;-\frac{am \Lambda M}{78732} \left(1-9 \Lambda  M^2\right) \left(9 \Lambda  M^2 \left(27 \Lambda  M^2 \left(3 \Lambda  M^2 \left(9 \Lambda 
   M^2 \left(373626 \Lambda  M^2-158551\right)+8158\right)+67040\right)-39581\right)-78904\right).\nonumber\\
\end{align}
\end{widetext}

The axial and polar gravitational QNMs are clearly isospectral to $O(L^{-3})$, with any difference between the two only becoming apparent in the case that both $a\neq0$ \textit{and} $\Lambda\neq0$. 

\section{QNM frequency tables}\label{tables}

\begin{table*}
\caption{Comparison of the $n=0$ scalar QNM frequencies calculated by 6th-order WKB \cite{Zhidenko:2003wq} and analytical expansion techniques for varying $\Lambda$ with $a=0$.}
\label{sdstablescalar}
\begin{ruledtabular}
\begin{tabular}{cccccccc}
\multicolumn{2}{l}{} &
\multicolumn{2}{c}{WKB}&
\multicolumn{2}{c}{L-expansion}&
\multicolumn{2}{c}{\% error}\\
 & $\Lambda$ & Re($M\omega$) & -Im($M\omega$)  & Re($M\omega$) & -Im($M\omega$) & Re($M\omega$) & -Im($M\omega$)
\\
\hline
\\
$\ell=1$ &0.00 &   0.2929 & 0.0978   & 0.292924 & 0.097649  & 0.00813958   & -0.155475  \\
%
%
	&0.04 & 0.2247 & 0.0821   & 0.224658 &  0.0820698 &  -0.0185787  & -0.0367874 \\
%
%
	&0.08 & 0.1404 & 0.0542  & 0.140395   & 0.0540001  & -0.00327199   &  -0.368871   \\
%
%
	&0.10 & 0.08156 & 0.03121  & 0.0816053   & 0.0312282  & 0.0555735   &  0.0581699   \\
%
%
\hline
\\
$\ell=2$ &0.00 &   0.48364 & 0.09677   & 0.483643 & 0.097649  & 0.000697739   & --0.0118938  \\
%
%
	&0.04 & 0.38078 & 0.07876   & 0.380783 &  0.086474 &  0.000776734  & -0.00117547 \\
%
%
	&0.08 & 0.24747 & 0.05197  & 0.24747   & 0.071116  & -0.000193423   &  -0.127102   \\
%
%
	&0.10 & 0.14661 & 0.03069  & 0.14661   & 0.071116  & 0.000330504   &  -0.010473   \\
%
%
\end{tabular}
\end{ruledtabular}
\end{table*}

\begin{table*}
\caption{Comparison of the $n=0$ electromagnetic QNM frequencies calculated by 6th-order WKB \cite{Zhidenko:2003wq} and analytical expansion techniques for varying $\Lambda$ with $a=0$.}
\label{sdstablevector}
\begin{ruledtabular}
\begin{tabular}{cccccccc}
\multicolumn{2}{l}{} &
\multicolumn{2}{c}{WKB}&
\multicolumn{2}{c}{L-expansion}&
\multicolumn{2}{c}{\% error}\\
 & $\Lambda$ & Re($M\omega$) & -Im($M\omega$)  & Re($M\omega$) & -Im($M\omega$) & Re($M\omega$) & -Im($M\omega$)
\\
\hline
\\
$\ell=1$ &0.00 &   0.2482 & 0.0926   & 0.248232 & 0.0924786  & 0.0130114   & -0.131053  \\
%
%
	&0.04 & 0.2006 & 0.0748   & 0.200597 &  0.0747225 &  -0.00135515  & -0.103609 \\
%
%
	&0.08 & 0.1339 & 0.0502  & 0.13392   & 0.0501975  & 0.0152628   &  -0.00490953   \\
%
%
	&0.10 & 0.08035 & 0.03028  & 0.0803542   & 0.0302716  & 0.00522626   &  -0.0277486   \\
%
%
\hline
\\
$\ell=2$ &0.00 &   0.45759 & 0.09501   & 0.457594 & 0.0950042  &0.000971495   & -0.00615498 \\
%
%
	&0.04 & 0.36723 & 0.07624   & 0.367228 & 0.0762387 &  -0.000646795  & -0.00176728 \\
%
%
	&0.08 & 0.24365 & 0.05067  & 0.243643 & 0.0506728 & -0.00280149   &  0.00554818   \\
%
%
	&0.10 & 0.14582 & 0.03037  & 0.145818 & 0.0303741  & -0.00149621   & 0.0133652  \\
%
%
\end{tabular}
\end{ruledtabular}
\end{table*}

\begin{table*}
\caption{Comparison of the $n=0$ gravitational QNM frequencies calculated by 6th-order WKB \cite{Zhidenko:2003wq} and analytical expansion techniques for varying $\Lambda$ with $a=0$.}
\label{sdstableaxial}
\begin{ruledtabular}
\begin{tabular}{cccccccc}
\multicolumn{2}{l}{} &
\multicolumn{2}{c}{WKB}&
\multicolumn{2}{c}{L-expansion}&
\multicolumn{2}{c}{\% error}\\
 & $\Lambda$ & Re($M\omega$) & -Im($M\omega$)  & Re($M\omega$) & -Im($M\omega$) & Re($M\omega$) & -Im($M\omega$)
\\
\hline
\\
$\ell=2$ &0.00 &   0.3736 & 0.0889   & 0.373642 & 0.0887156  & 0.0113413   & -0.207368  \\
%
%
	&0.04 & 0.2989 & 0.0733   & 0.299056 &  0.0731413 &  0.0521161  & -0.216536 \\
%
%
	&0.08 & 0.1975 & 0.0499  & 0.197732   & 0.0498251  & 0.117397   &  -0.150176   \\
%
%
	&0.10 & 0.11792 & 0.03021  & 0.118107   & 0.0301989  & 0.158985   &  -0.0367963   \\
%
%
\hline
\\
$\ell=3$ &0.00 &   0.599443 & 0.092703   & 0.599439 & 0.0926902  &-0.000635737   & -0.0138235 \\
%
%
	&0.04 & 0.480058 & 0.075146   & 0.48007 & 0.0751385 &  0.00240952  & -0.0100284 \\
%
%
	&0.08 & 0.317805 & 0.050382  & 0.317824 & 0.0503795 & 0.00608041   &  0.00501461   \\
%
%
	&0.10 & 0.189994 & 0.030314  & 0.190009 & 0.030314  & 0.00806499   & 0.000105339  \\
%
%
\end{tabular}
\end{ruledtabular}
\end{table*}

\begin{table*}
\caption{Comparison of the $n=0$ massless scalar QNM frequencies as calculated by Leaver's continued fraction method in \cite{Berti:2009kk} and analytical expansion techniques for varying $a$ with $\Lambda=0$.}
\label{kerrtablescalar}
\begin{ruledtabular}
\begin{tabular}{cccccccc}
\multicolumn{2}{l}{} &
\multicolumn{2}{c}{Leaver}&
\multicolumn{2}{c}{L-expansion}&
\multicolumn{2}{c}{\% error}\\
 & $a$ & Re($M\omega$) & -Im($M\omega$)  & Re($M\omega$) & -Im($M\omega$) & Re($M\omega$) & -Im($M\omega$)
\\
\hline
\\
$\ell=m=1$ &0.05 &   0.296889 & 0.0976242   & 0.296781 & 0.0976311  & -0.0363486   & 0.00706991  \\
	&0.10 & 0.301045 & 0.0975472   & 0.300639 &  0.0976143 &  -0.134768  & 0.068833 \\
	&0.15 & 0.305421 & 0.0974231  & 0.304496   & 0.0975975  & -0.302922   &  0.179031   \\
	&0.20 & 0.310043 & 0.097245  & 0.308354   & 0.0975807  & -0.544711   &  0.345159   \\
\hline
\\
$\ell=m=2$ &0.05 &   0.49136 & 0.0967329   & 0.491168 & 0.0967513  &-0.0390091   & 0.0190342 \\
	&0.10 & 0.499482 & 0.096666   & 0.498692 & 0.0967442 &  -0.158139  & 0.0808871 \\
	&0.15 & 0.508053 & 0.0965516  & 0.506217 & 0.096737 & -0.361464   &  0.192037   \\
	&0.20 & 0.517121 & 0.0963822  & 0.513741 & 0.0967299  & -0.653486   & 0.360803  \\
\end{tabular}
\end{ruledtabular}
\end{table*}

\begin{table*}
\caption{Comparison of the $n=0$ electromagnetic QNM frequencies as calculated by Leaver's continued fraction method in \cite{Berti:2009kk} and analytical expansion techniques for varying $a$ with $\Lambda=0$.}
\label{kerrtableelec}
\begin{ruledtabular}
\begin{tabular}{cccccccc}
\multicolumn{2}{l}{} &
\multicolumn{2}{c}{Leaver}&
\multicolumn{2}{c}{L-expansion}&
\multicolumn{2}{c}{\% error}\\
 & $a$ & Re($M\omega$) & -Im($M\omega$)  & Re($M\omega$) & -Im($M\omega$) & Re($M\omega$) & -Im($M\omega$)
\\
\hline
\\
$\ell=m=1$ &0.05 &   0.251642 & 0.092286   & 0.251517 & 0.0923056  & -0.0499051   & -0.0212227  \\
	&0.10 & 0.255214 & 0.0920427   & 0.254081 &  0.0921325 &  -0.161837  & 0.0975817 \\
	&0.15 & 0.258999 & 0.0917527  & 0.258086   & 0.0919594  & -0.352633   &  0.22524   \\
	&0.20 & 0.26302 & 0.0914101  & 0.26137   & 0.0917863  & -0.627229   &  0.411579   \\
\hline
\\
$\ell=m=2$ &0.05 &   0.464904 & 0.0949194   & 0.464718 & 0.094939  &-0.0401393   & 0.0205991 \\
	&0.10 & 0.472609 & 0.0947922   & 0.471841 & 0.0948738 &  -0.162406  & 0.0860831 \\
	&0.15 & 0.48075 & 0.0946166  & 0.478965 & 0.0948087 & -0.371317   &  0.203022   \\
	&0.20 & 0.489375 & 0.094385  & 0.486088 & 0.0947435  & -0.671741   & 0.379827  \\
\end{tabular}
\end{ruledtabular}
\end{table*}

\begin{table*}
\caption{Comparison of the $n=0$ gravitational QNM frequencies as calculated by Leaver's continued fraction method in \cite{Berti:2009kk} and analytical expansion techniques for varying $a$ with $\Lambda=0$.}
\label{kerrtablegrav}
\begin{ruledtabular}
\begin{tabular}{cccccccc}
\multicolumn{2}{l}{} &
\multicolumn{2}{c}{Leaver}&
\multicolumn{2}{c}{L-expansion}&
\multicolumn{2}{c}{\% error}\\
 & $a$ & Re($M\omega$) & -Im($M\omega$)  & Re($M\omega$) & -Im($M\omega$) & Re($M\omega$) & -Im($M\omega$)
\\
\hline
\\
$\ell=m=2$ &0.05 &   0.380146 & 0.0888489   & 0.379911 & 0.0885439  & 0.0616187   & -0.343309  \\
	&0.10 & 0.387018 & 0.0887057   & 0.386180 &  0.0883721 &  -0.216332  & -0.376072 \\
	&0.15 & 0.394333 & 0.0885283  & 0.392449   & 0.0882003  & -0.4777   &  -0.370435   \\
	&0.20 & 0.402145 & 0.0883112  & 0.398718   & 0.0880286  & -0.852205   &  -0.320018   \\
\hline
\\
$\ell=m=3$ &0.05 &   0.609823 & 0.0925869   & 0.609540 & 0.0925840  &-0.0463903   & -0.00312067 \\
	&0.10 & 0.620796 & 0.0924305   & 0.619642 & 0.0924778 &  -0.185874  & 0.0511351 \\
	&0.15 & 0.632425 & 0.0922281  & 0.629743 & 0.0923715 & -0.424063   &  0.155544   \\
	&0.20 & 0.644787 & 0.0919726  & 0.639844 & 0.0922653  & -0.76657   & 0.318254  \\
\end{tabular}
\end{ruledtabular}
\end{table*}

\clearpage

\bibliography{RefModifiedGravity}

\end{document}